\documentclass[12pt]{iopart}

\usepackage{graphicx}
\usepackage{epsfig}

\begin{document}

\title{Nonlinear Jaynes-Cummings model for two interacting two-level atoms}
\author{O de los Santos-S\'anchez, C Gonz\'alez-Guti\'errez and J R\'ecamier} 

\address{Instituto de Ciencias F\'isicas, Universidad Nacional Aut\'onoma de M\'exico, 
62210 Cuernavaca, Morelos, M\'exico}
 
\eads{\mailto{octavio.desantos@gmail.com, carlosag@fis.unam.mx}}

\begin{abstract}
In this work we examine a nonlinear version of the Jaynes-Cummings model for two identical two-level atoms allowing for Ising-like and dipole-dipole interplays between them. The model is said to be  nonlinear in the sense that it can incorporate both a general intensity-dependent interaction between the atomic system and the cavity field and/or the presence of a nonlinear medium inside the cavity. As an example, we consider a particular type of atom-field coupling based upon the so-called Buck-Sukumar model and a lossless Kerr-like cavity. We describe the possible effects of such features on the evolution of some quantities of current interest, such as atomic excitation, purity, concurrence, the entropy of the field and the evolution of the latter in phase space. 
\end{abstract}

\maketitle

\section{Introduction}
The standard Jaynes-Cummings model (JCM) is usually referred to as the simplest fully quantized light-matter interaction scheme involving just one atom and one mode of the radiation field. Since its introduction in 1963 \cite{jaynes}, the model has attracted a great deal of theoretical and experimental interest in the fields of laser physics and quantum optics \cite{shore}, and this trend is due in part to its apparent simplicity and, more importantly, to its striking predictions about the dynamical properties of the subsystems involved with which the quantum optics community is now familiar, as for example the well-known phenomenon of {\it collpases} and {\it revivals} for the atomic excitation understood in terms of the grainy statistics of the radiation field \cite{eberly}. \\

The JCM has become the source of inspiration for a wide variety of generalizations dealing with more general and/or realistic circumstances. The majority of them focused primarily upon multi-photon transitions and/or multimode fields \cite{parker,parkins,jandp}; engineering nonlinear atom-field couplings, like the Buck-Sukumar and Kochetov models \cite{sukumar,kochetov} (including some very  recent  improvements and extensions of the former with or without applying the rotating wave approximation  \cite{cordeiro,penna}), one-atom JC models involving two-photon interaction with intensity-dependence both in the atom-field coupling and the detuning \cite{napoli1,napoli2}, or adding nonlinear Kerr-like media \cite{buzek,joshipuri,guido,cordero}; interacting or noninteracting two two-level atoms \cite{deng,joshi,zhang,chilingaryan}; or even more complex systems involving a large group of $N$ two-level (or multi-level) atoms in the same cavity, such as the so-called Tavis-Cummings model (TCM) \cite{tavis1,tavis2}, to mention some examples. And still more recently,  renewed attention has been paid to quantum decoherence and  entanglement properties of light-matter interaction models \`a la Jaynes-Cummings whose central system is composed of two or three two-level atoms (also called within the jargon of the quantum information framework as two- or three-qubit systems) resonantly coupled with a cavity field prepared in a single number state and also with each other through dipole-dipole and Ising-like interactions \cite{torres1,pineda} (incidentally, Spin-Spin interactions such as these or similar have also been the theme of current interest in a manifold of areas such as optical lattices \cite{sorensen}; systems with trapped ions \cite{porras} and microcavities \cite{hartmann}; and, even in the context of nearly localized and dipolarly coupled two identical molecules \cite{napoli3,napoli4}, where, under certain conditions, from an algebraic-structure point of view, intramolecular coupling models \`a la Jaynes-Cummings emerge); moreover, along this line an interesting application based on the resonant two-atom JCM has been proposed with the aim of implementing novel protocols for unambiguous Bell state discrimination for two qubits \cite{torres2}. \\

This work is in keeping with the aforementioned spirit of putting forward another extension of the Jaynes-Cummings model, that is, a nonlinear version of it for the description of a qubit system composed of two identical two-level atoms that interact with each other via dipole-dipole and Ising-like interaction and with a one-mode cavity field being in a coherent superposition of number states. The nonlinear character of the model is construed  in a twofold sense: i) the interaction between the atomic system and the field is considered to be nonlinear in a way such that it hinges on the number of photons of the latter, and ii) the said atomic system can be allowed to be embedded in a kind of nonlinear medium inside the cavity. In considering a Buck-Sukumar nonlinear coupling \cite{sukumar} between the atoms and the field and a Kerr-like medium within the cavity as particular cases, a full algebraic solution of the problem is provided and the consequences of incorporating such nonlinearities  are illustrated by exploring, in the resonant quantum dynamics, some quantities of current physical interest, such as population inversion, purity, and concurrence, as far as the atomic system is concerned, and the time behavior of the field itself is also investigated in terms of its entropy and its image on phase space. It is worth commenting that the set of results reported here, regarding the aforesaid nonlinear coupling scheme, may also be of some relevance in the light of novel experimental and theoretical research on optical simulation of the Jaynes-Cummings and Rabi models in arrays of coupled photonic waveguides \cite{crespi,blas1,blas2}, as well as in current designs of architectures intended for quantum computation and communication based on cavity quantum electrodynamics (CQED) studies for solid-state superconducting electrical circuits \cite{blais}. Furthermore, we should stress the fact that the present treatment is not only valid for interactions between quantized fields and atoms, but also in the context of trapped ions interacting with lasers (classical) fields where a number of phenomena known from nonlinear optics, including Kerr-like nonlinearities, can be easily produced (see, for instance, Refs. \cite{wallentowitz,matos,manko,blatt,moya}). So, from our point of view, the foregoing contributions may open up the possibility of realizing Hamiltonian models such as the  one we are considering in physically different but algebraically equivalent contexts. \\

The content of the paper is the following. In section 2 the Hamiltonian model and its general solution for two  Ising- and dipole-dipole interacting and identical two-level atoms are introduced. Section 3, which in turn is divided into three subsections, is thoroughly devoted to the discussion of some properties of the atomic system in terms of its population, purity and concurrence dynamics; all these features are explored by taking into consideration each nonlinear contribution of the model. In section 4, the effect of such nonlinearities on the evolution of the cavity field is explored based upon its entropy and its  phase space picture via the Q-function.  And finally, in section 5 some conclusions are given. 

\section{The model and its solution}
Let the Hamiltonian model of the system under study be encoded as follows:   
\begin{equation}
H = H_{F}+H_{A}+H_{FA}+H_{AA},
\label{eq:totalham}
\end{equation}
where the constituent Hamiltonians are explicitly given, in the Schr\"odinger picture, and under the rotating-wave approximation, by
\begin{eqnarray}
H_{F} & = & \hbar \omega_{0} \hat{n}h(\hat{n}), \label{eq:hamfield}\\
H_{A} & = & \frac{\hbar \omega}{2} (\sigma_{Z}^{(1)}+\sigma_{Z}^{(2)}), \\
H_{FA} & = & \hbar g(\sigma_{+}^{(1)}\hat{a}f(\hat{n})+\sigma_{-}^{(1)}f(\hat{n})\hat{a}^{\dagger})+\hbar g(\sigma_{+}^{(2)}\hat{a}f(\hat{n})+\sigma_{-}^{(2)}f(\hat{n})\hat{a}^{\dagger}),\\
H_{AA} & = & 2\hbar \kappa (\sigma_{-}^{(1)}\sigma_{+}^{(2)}+\sigma_{+}^{(1)}\sigma_{-}^{(2)})+\hbar J\sigma_{Z}^{(1)}\sigma_{Z}^{(2)}. \label{eq:haa}
\end{eqnarray}
Here, $H_{F}$ and $H_{A}$ are the energy operators for the field and atoms, respectively, the coupling between the atomic system and the radiation field is described by $H_{FA}$, and $H_{AA}$ is the contribution to the total system of the atom-atom interaction mentioned above. The field mode frequency is $\omega_{0}$, $\omega$ is the atomic transition frequency, $g$ is the coupling constant (which is taken to be the same for both atoms), and $J$ and $\kappa$ are, respectively, the Ising and dipole-dipole parameters. As usual, $\hat{a}$ ($\hat{a}^{\dagger}$) is the photon annihilation (creation) operator satisfying $[\hat{a},\hat{a}^{\dagger}]=1$, and, on the other side, $\hat{\sigma}_{Z}^{(1)}$ ($\hat{\sigma_{Z}}^{(2)}$), $\hat{\sigma}_{\pm}^{(1)}$ ($\hat{\sigma}_{\pm}^{(2)}$) are the standard atomic two-level transition operators of the respective atoms. Finally, $h(\hat{n})$ and $f(\hat{n})$, with $\hat{n}=\hat{a}^{\dagger}\hat{a}$, are photon-number-dependent functions that represent the aforesaid nonlinear character of the whole system. 

At this point, we note in passing that although we shall be working with parameters such that the rotating-wave approximation is considered to be valid, we also have to be careful with the field intensities reflected upon the  functions $f(\hat{n})$ and $h(\hat{n})$, always taking into account that the conditions $g\langle f(\hat{n}) \rangle \ll \omega_{0}\langle h(\hat{n})\rangle$ and  $g\langle f(\hat{n}) \rangle \ll \omega$ are fulfilled and that the effective detuning $ \omega-\omega_{0}\langle  (\hat{n}+1)h(\hat{n}+1)- \hat{n}h(\hat{n})\rangle$ still allows us to work within the two-level atom approximation. The origin of and the role played by such an intensity-dependent detuning can readily be seen if we reframed our system under the transformation $e^{i(H_{F}+H_{A})t/\hbar} H e^{-i(H_{F}+H_{A})t/\hbar}$ to get an atom-field coupling structure of the form $\sim \hbar g e^{-i\delta_{\hat{n}}t} \sigma_{+}^{(j)}\hat{a}f(\hat{n})$, with the fitting identification $\delta_{\hat{n}}=\omega-\omega_{0} \left \{ (\hat{n}+1)h(\hat{n}+1)-\hat{n}h(\hat{n}) \right \}$ of the detuning we are referring to. The intensity-dependent part of the latter is construed as the frequency separation between adjacent energy levels of a nonlinear field (i.e., a nonlinear oscillator, viewed as a whole) whose spectrum is clearly nonequidistant by virtue of $h(\hat{n})$ in the definition of the field Hamiltonian $H_{F}$. \\

Here, we find it convenient to choose the framework generated by the unitary operator $V = \exp \left [-i\omega_{0}t\left (\hat{n}+\frac{\sigma_{Z}^{(1)}}{2}+\frac{\sigma_{Z}^{(2)}}{2}\right)\right]$ so that one can recast Hamiltonian (\ref{eq:totalham}), on inserting it into the transformation
\begin{equation}
H_{I} = V^{\dagger}H V-i\hbar V^{\dagger} \frac{dV}{dt},
\end{equation}
as follows
\begin{eqnarray}
H_{I} & = & \hbar \omega_{0} \hat{n} (h(\hat{n})-1)+\frac{\hbar \delta}{2}(\sigma_{Z}^{(1)}+\sigma_{Z}^{(2)})+\hbar (Df(\hat{n})\hat{a}^{\dagger}+D^{\dagger}\hat{a}f(\hat{n})) \nonumber \\
& & +2\hbar \kappa (\sigma_{-}^{(1)}\sigma_{+}^{(2)}+\sigma_{+}^{(1)}\sigma_{-}^{(2)})+\hbar J\sigma_{Z}^{(1)}\sigma_{Z}^{(2)},
\label{eq:totalhamtrans}
\end{eqnarray}
where we have labeled $\delta = \omega-\omega_{0}$ and $D=g(\sigma_{-}^{(1)}+\sigma_{-}^{(2)})$. It is well-known that the dynamics of a system such as this involving a single-photon process where the number of excitations is a conserved quantity (as can be verified since the excitation number operator $\hat{N}=\hat{n}+\frac{1}{2}(\sigma^{(1)}+\sigma^{(2)})$ commutes with Hamiltonian (\ref{eq:totalhamtrans})) the set of states defined by $\{ |e,e,n\rangle , |e,g,n+1 \rangle, |g,e,n+1\rangle, |g,g,n+2\rangle \}$ may be employed as a basis to diagonalize the Hamiltonian under study (see, for instance, Refs. \cite{deng,joshi,zhang,torres1}). However, one can see that it is also possible to employ the following fitting basis:
\begin{eqnarray}
|\phi_{1}^{(n)} \rangle & = & |e,e,n\rangle , \label{eq:state1} \\
|\phi_{2}^{(n)} \rangle & = & \frac{1}{\sqrt{2}}(|e,g,n+1\rangle +|g,e,n+1\rangle), \label{eq:state2} \\
|\phi_{3}^{(n)} \rangle & = & |g,g,n+2 \rangle, \label{eq:state3}
\end{eqnarray}
where the use of the symmetric combination of $|e,g,n+1\rangle$ and $|g,e,n+1\rangle $ turns out to be  convenient  in the sense that it will allows us to simplify subsequent calculations (such a symmetric state, needless to say, is useful only in the particular case when the two atoms are identical). So, in this basis the representation of $H_{I}$ is a $3\times 3$ matrix with elements $H_{ij}^{(n)}=\langle \phi_{i}^{(n)}|H_{I}| \phi_{j}^{(n)} \rangle$ for a given number of photons $n$. In units where $\hbar=1$ we get: 

\begin{equation}
H_{I} \doteq \left(\begin{array}{ccc} \omega_{0}F_{n,0}+ \delta+J & \sqrt{2} g f_{n+1} & 0 \\ \sqrt{2} g f_{n+1} &  \omega_{0}F_{n,1}- J+2 \kappa & \sqrt{2} g f_{n+2} \\ 0 & \sqrt{2} gf_{n+2} & \omega_{0}F_{n,2}-\delta+J \end{array} \right),
\end{equation}
where, for the sake of simplicity, we have utilized the shorthand notation $f_{n+1}=f(n+1)\sqrt{n+1}$, $F_{n,i}=(n+i)(h(n+i)-1)$, with $i=0,1,2$.\\

For all the above initial considerations, we now proceed to solve the eigenvalue equation
\begin{equation}
H_{I}|u_{j}^{(n)} \rangle = E_{j}^{(n)} |u_{j}^{(n)} \rangle,
\end{equation}
for $j=1,2,3$. Tackling this problem entails solving a cubic polynomial equation whose roots are well-known. So, it follows straightforwardly from Cardano's formulae that the sought eigenvalues $E_{j}^{(n)}$ are given by
\begin{equation}
E_{j}^{(n)}  =  -\frac{1}{3}\beta_{n}+ 2\sqrt{-Q_{n}}\cos \left(\frac{\theta_{n}+2(j-1)\pi}{3}\right),
\end{equation}
with
\begin{eqnarray}
Q_{n} & = &\frac{3\gamma_{n} - \beta^{2}_{n}}{9}, \label{eq:q}\\
R_{n} & = & \frac{9\beta_{n} \gamma_{n}-27 \eta_{n}-2\beta^{3}_{n}}{54}, \label{eq:r}\\
\theta_{n} & = & \cos^{-1}\left ( \frac{R_{n}}{\sqrt{-Q^{3}_{n}}} \right ). \label{eq:theta}
\end{eqnarray}
Such eigenvalues are all real provided that $Q^{3}_{n}+R^{2}_{n} <0$, with $\beta_{n}$, $\gamma_{n}$ and $\eta_{n}$ being, respectively,
\begin{eqnarray}
\beta_{n} & = & - (\omega_{0}F_{n}+J+2\kappa), \\
\gamma_{n} & = &  (\omega_{0}F_{n,1}-J+2\kappa)(\omega_{0}(F_{n,0}+F_{n,2})+2J) \nonumber \\
& & +(\omega_{0}F_{n,0}+\delta+J)(\omega_{0}F_{n,2}-\delta+J)-2 g^{2}\Delta_{n}^{+},\\
\eta_{n} & = & -(\omega_{0}F_{n,2}-\delta+J)(\omega_{0}F_{n,0}+\delta+J)(\omega_{0}F_{n,1}+2\kappa-J) \nonumber \\
& & +2 g^{2}(\omega_{0}G_{n}+\delta \Delta_{n}^{-}+J\Delta_{n}^{+}),
\end{eqnarray}
where we have set the photon-number dependent functions:
\begin{eqnarray}
F_{n} & = & F_{n,0}+F_{n,1}+F_{n,2} ,\\
\Delta_{n}^{\pm} & = & f^{2}_{n+2} \pm f^{2}_{n+1}, \label{eq:deltan}\\
G_{n} & = & F_{n,0}f^{2}_{n+2}+F_{n,2}f^{2}_{n+1}.
\end{eqnarray}
So, we let the corresponding set of normalized eigenvectors be written as
\begin{eqnarray}
\fl |u_{j}^{(n)} \rangle  & = &   \frac{1}{N_{j}^{(n)}} \Bigg ( H_{12}^{(n)}H_{23}^{(n)} |\phi_{1}^{(n)}\rangle+H_{23}^{(n)} (E_{j}^{(n)}-H_{11}^{(n)}) |\phi_{2}^{(n)}\rangle \nonumber \\
\fl & & + \left((E_{j}^{(n)}-H_{22}^{(n)})(E_{j}^{(n)}-H_{11}^{(n)})-H_{12}^{(n) 2}\right) |\phi_{3}^{(n)}\rangle \Bigg),
\label{eq:vector}
\end{eqnarray}
for $j=1,2,3$, and where the normalization factor $N_{j}^{(n)}$ reads
\begin{equation}
\fl N_{j}^{(n)} = \left( H_{12}^{(n) 2} H_{ 23}^{(n) 2}+\left(E_{j}^{(n)}-H_{11}^{(n)}\right)^{2}H_{23}^{(n)2}+ \left (\left (E_{j}^{(n)}-H_{22}^{(n)}\right)\left (E_{j}^{(n)}-H_{11}^{(n)}\right )-H_{12}^{(n) 2}\right)^{2} \right)^{1/2}.
\end{equation}
In order to further simplify subsequent algebraic manipulations we let (\ref{eq:vector}) be recast as
\begin{equation}
|u_{j}^{(n)} \rangle = \sum_{k=1}^{3} C_{jk}^{(n)} |\phi_{k}^{(n)} \rangle,
\label{eq:basis1}
\end{equation}
with the following correspondences 
\begin{eqnarray}
C_{j,1}^{(n)} & = & \frac{H_{12}^{(n)}H_{23}^{(n)}}{N_{j}^{(n)}},  \label{eq:cj1}\\
C_{j,2}^{(n)} & = & \frac{H_{23}^{(n)} \left (E_{j}^{(n)}-H_{11}^{(n)}\right)}{N_{j}^{(n)}},  \label{eq:cj2} \\
C_{j,3}^{(n)} & = & \frac{\left (E_{j}^{(n)}-H_{11}^{(n)}\right)\left(E_{j}^{(n)}-H_{22}^{(n)}\right)-H_{12}^{(n) 2}}{N_{j}^{(n)}}, \label{eq:cj3}
\end{eqnarray}
from which one can also deduce that  
\begin{equation}
|\phi_{j}^{(n)} \rangle = \sum_{k=1}^{3} C_{kj}^{(n)} |u_{k}^{(n)} \rangle.
\label{eq:basis2}
\end{equation}
This set of results will be our starting point for examining some properties of physical interest concerning the evolution of either the atomic system or the cavity field, such as the population inversion,  entanglement dynamics based on the purity and concurrence features, and the evolution of the field on phase space with the help of the Q-function representation.   
\section{Evolution of the atomic system: atomic excitation, purity and concurrence}
\subsection{Population inversion}
Let us first examine the effect of dipole-dipole and Ising-like interactions on the dynamics of the atomic level occupation. The time evolution of this physical feature can be assessed by considering the inversion operator \cite{zhang}
\begin{equation}
D_{Z}= \frac{1}{2}(\sigma_{z}^{(1)} + \sigma_{z}^{(2)}),
\label{eq:opdz}
\end{equation}
together with the use of the wave function 
\begin{equation}
|\Phi(t) \rangle = e^{-iH_{I}t} |\Phi (0) \rangle,
\end{equation}
where $|\Phi (0) \rangle$ is the initial state of the whole system. For the time being, let the initial state be in a way such that the atoms are both in their corresponding excited state and the cavity field in a coherent superposition of number states, i.e., 
\begin{equation}
|\Phi (0) \rangle = \sum_{n=0}^{\infty} A_{n}|e,e,n \rangle=\sum_{n}^{\infty} A_{n}|\phi_{1}^{(n)} \rangle ,
\end{equation}
which, by utilizing (\ref{eq:basis2}), can be rewritten as
\begin{equation}
|\Phi (0) \rangle = \sum_{n=0}^{\infty}\sum_{k=1}^{3} A_{n} C_{k1}^{(n)} |u_{k}^{(n)} \rangle,
\end{equation}
so that
\begin{eqnarray}
|\Phi(t) \rangle & = & \sum_{n=0}^{\infty}\sum_{k=1}^{3} A_{n} C_{k1}^{(n)} e^{-iE_{k}^{(n)}t} |u_{k}^{(n)} \rangle, \nonumber \\
& = & \sum_{n=0}^{\infty}\sum_{k,j=1}^{3} A_{n}C_{j1}^{(n)}C_{jk}^{(n)}e^{-iE_{j}^{(n)}t}|\phi_{k}^{(n)} \rangle, \nonumber \\
& = & \sum_{n=0}^{\infty}\sum_{k=1}^{3} A_{n} \mathcal{D}_{k}^{(n)}(t)|\phi_{k}^{(n)} \rangle,
\label{eq:wavefunc}
\end{eqnarray}
where we have set the time-dependent coefficient 
\begin{equation}
\mathcal{D}_{k}^{(n)}(t) = \sum_{j=1}^{3} C^{(n)}_{j1}C^{(n)}_{jk} e^{-iE_{j}^{(n)}t}.
\label{eq:coefficient}
\end{equation}
By taking the expectation value of (\ref{eq:opdz}) with the help of (\ref{eq:wavefunc}), and after making some rearrangements, we arrive at the desired result for the atomic inversion
\begin{equation}
\fl \langle \Phi(t)|D_{Z}|\Phi (t) \rangle  =   \sum_{n=0}^{\infty}P_{n} \left( \sum_{j=1}^{3} \Lambda_{jj}^{(n)} +2  \left \{ \Lambda_{21}^{(n)} \cos \left (\Omega_{21}^{(n)}t\right ) +\Lambda_{31}^{(n)} \cos \left ( \Omega_{31}^{(n)} t \right ) +\Lambda_{23}^{(n)} \cos \left (\Omega_{23}^{(n)} t \right) \right \} \right), 
\label{eq:inversion}
\end{equation}
in which $P_{n} = |A_{n}|^{2}$ represents the initial photon distribution function of the field, 
\begin{eqnarray}
\Omega_{21}^{(n)} & = & E_{1}^{(n)}-E_{2}^{(n)} =\sqrt{-3Q_{n}}  \left( \sqrt{3} \cos \left( \frac{\theta_{n}}{3} \right) +\sin \left( \frac{\theta_{n}}{3} \right) \right), \label{eq:rabi1} \\
\Omega_{31}^{(n)} & = & E_{1}^{(n)}-E_{3}^{(n)} =  \sqrt{-3Q_{n}}  \left(  \sqrt{3} \cos \left( \frac{\theta_{n}}{3} \right) -\sin \left( \frac{\theta_{n}}{3} \right) \right),    \label{eq:rabi2} \\
\Omega_{23}^{(n)} & = & E_{3}^{(n)}-E_{2}^{(n)}= 2 \sqrt{-3Q_{n}} \sin \left( \frac{\theta_{n}}{3} \right), \label{eq:rabi3}
\end{eqnarray}
with $Q_{n}$ and $\theta_{n}$ being given by (\ref{eq:q}) and (\ref{eq:theta}), respectively. And, following a similar notation as in Ref. \cite{zhang}, the weighting amplitudes $\Lambda_{jk}$'s take the form
\begin{equation}
\Lambda_{jk}^{(n)} = C_{j1}^{(n)}C_{k1}^{(n)}\left (C_{j1}^{(n)}C_{k1}^{(n)}-C_{j3}^{(n)}C_{k3}^{(n)}\right), \label{eq:amplitudes}
\end{equation}
where, in turn, the $C_{i1}$ and $C_{i3}$ are given by (\ref{eq:cj1}) and (\ref{eq:cj3}). \\

From Eq. (\ref{eq:inversion}) one can readily see that, as opposed to the well-known one-atom model, the dynamics of atomic inversion consists, save for a constant term, essentially of a superposition of three different oscillatory components, each giving rise to three different patterns of collapses and revivals and characterized by its own Rabi frequency (see Eqs. (\ref{eq:rabi1})-(\ref{eq:rabi2})). It is found that this set of frequencies satisfies, in general, the following relationships: 
\begin{equation}
\Omega_{21}^{(n)}  =  \Omega_{23}^{(n)}+\Omega_{31}^{(n)}, \\
\end{equation}
\begin{equation}
\frac{1}{3} \left( \Omega_{23}^{(n)}+2\Omega_{31}^{(n)} \right)^{2}+\Omega_{23}^{(n) 2}  =  4 |-3Q_{n}| ,
\end{equation}
where the vertical bars $|\cdot|$ denote the absolute value of the quantity involved and 
\begin{eqnarray}
\fl Q_{n} & = &  \frac{1}{9} \bigg \{ \omega_{0}^{2} \left[ F_{n,0}(F_{n,1}-F_{n,0}) +F_{n,1}(F_{n,2}-F_{n,1}) +F_{n,2}(F_{n,0}-F_{n,2}) \right] + \nonumber \\
\fl &  &+ 2\omega_{0}(J-\kappa)(2F_{n,1}-F_{n,0}-F_{n,2})+3\omega_{0} \delta (F_{n,2}-F_{n,0}) \nonumber \\
\fl & & -4(J-\kappa)^{2}-3\delta^{2}-6g^{2}\Delta_{n}^{+} \bigg \}.
\end{eqnarray}
A better picture of the evolution of this physical feature will be shown by the graphical representation of two particular cases: Firstly, we will consider a situation in which the atomic system is supposed to be immersed in a nonlinear Kerr-like medium within the cavity, so that the number operator function $h(n)$ in Eq. (\ref{eq:hamfield}) is chosen to have the form $h(n)=1+\frac{\chi}{\omega_{0}}n$, leading to the Hamiltonian of the field $H_{F} = \omega_{0}\hat{n}+ \chi \hat{n}^{2}$; here, $\chi$ is construed as an anharmonicity parameter associated with the dispersive part of the third-order nonlinearity of the medium within which the radiation field also evolves \cite{mandel}. [This is, incidentally, an approximate and justified model widely used in nonlinear optics \cite{agarwal,mancini} whenever the medium's time response is sufficiently small so that the medium itself may be regarded as being able to follow the field in an adiabatic manner \cite{agarwal}; an agreeable and thorough treatment of the subject had already been undertaken, for instance, by Bu\u zek and Jex \cite{buzek} in their work on the Jaynes-Cummings model with only one atom when the cavity is supposed to be filled with a Kerr-like medium]. Secondly, the influence of the interplay between the two atoms ($\kappa, J \neq 0$) upon the population dynamics is explored by taking into consideration a standard cavity, i.e., when $h(n)=1$ ($\chi=0$). Unless otherwise specified, in subsequent calculations the initial state of the field is considered to be a coherent state with a Poissonian statistiscal distribution  $P_{n}=e^{-\langle n \rangle}\frac{\langle n\rangle ^{n}}{n!}$ with $\langle n\rangle =10$ photons. 

\subsubsection{Case 1. Kerr-like cavity: $h(n)=1+\frac{\chi}{\omega_{0}}n $}

Merging the nonlinearity of the medium within the cavity with both the dipole-dipole and Ising-like atomic interactions leads to the following analytic expressions for Cardano's formulae $Q_{n}$ and $\theta_{n}$ (Eqs. (\ref{eq:q}) and (\ref{eq:theta})) in terms of the parameters $\chi$, $\kappa$, and $J$:
\begin{eqnarray}
\fl Q_{n} & = & -\frac{1}{9} \left \{ \left(\chi-2(\kappa-J)\right)^{2}+12  \chi^{2}(n+1)^{2}+6g^{2}\Delta_{n}^{+} \right \}, \label {eq:cardanosqt} \\
\fl \theta_{n} & = & \cos^{-1} \left \{ \frac{\left (\chi-2(\kappa-J) \right) \left[36 \chi^{2}(n+1)^{2}- \left (\chi-2(\kappa-J)\right )^{2}-9g^{2}\Delta_{n}^{+} \right]+54 g^{2}\chi (n+1)\Delta_{n}^{-}}{\left (\left(\chi-2(\kappa-J)\right)^{2}+12  \chi^{2}(n+1)^{2}+6g^{2}\Delta_{n}^{+} \right)^{3/2}} \right \}, \nonumber
\end{eqnarray}
where the number-dependent function $f(n)$ describing the nonlinear dipole coupling between the field and the atoms remains implicit into the frequency shifts $\Delta_{n}^{\pm}$ and  the cavity frequency is assumed to match the atomic transition frequency ($\delta=0$). It is worth highlighting that this set of equations can also be viewed as dependent on the difference $\kappa-J$ for a given $\chi$, instead of focusing separately on the values of $\kappa$ and $J$. Such a dependency is of course reflected upon the Rabi frequencies (\ref{eq:rabi1})-(\ref{eq:rabi3}) and the set of weighting amplitudes (\ref{eq:amplitudes}) by direct substitution of $Q_{n}$ and $\theta_{n}$ given above. It is clear then that the dipole-dipole and Ising interactions cancel each other when the equality $\kappa = J$ holds, as though there was a kind of trade-off between both atomic interactions. \\

\begin{figure}[h!]
\begin{center}
\includegraphics[width=7.5cm, height=7cm]{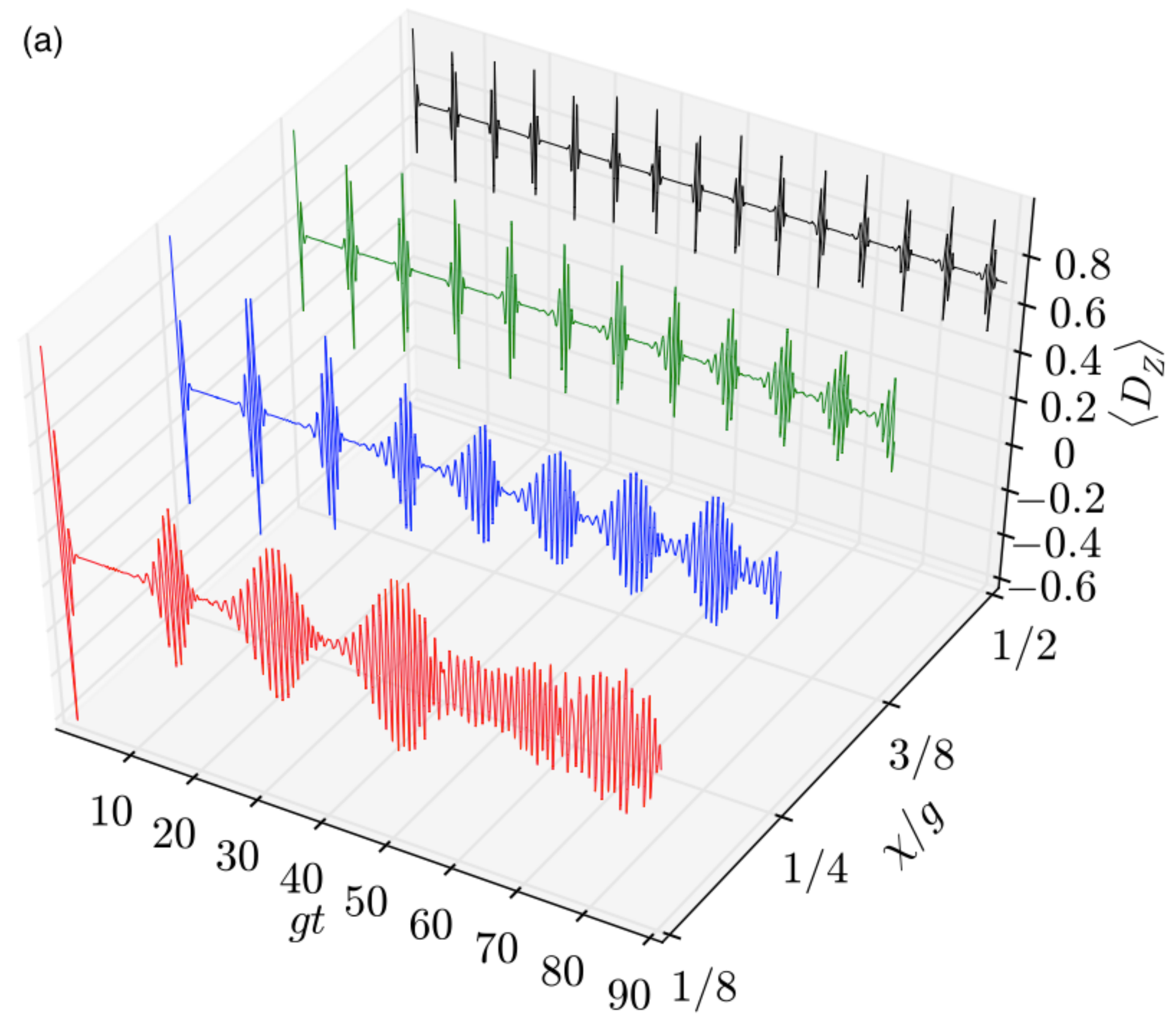} 
\includegraphics[width=7.5cm, height=7cm]{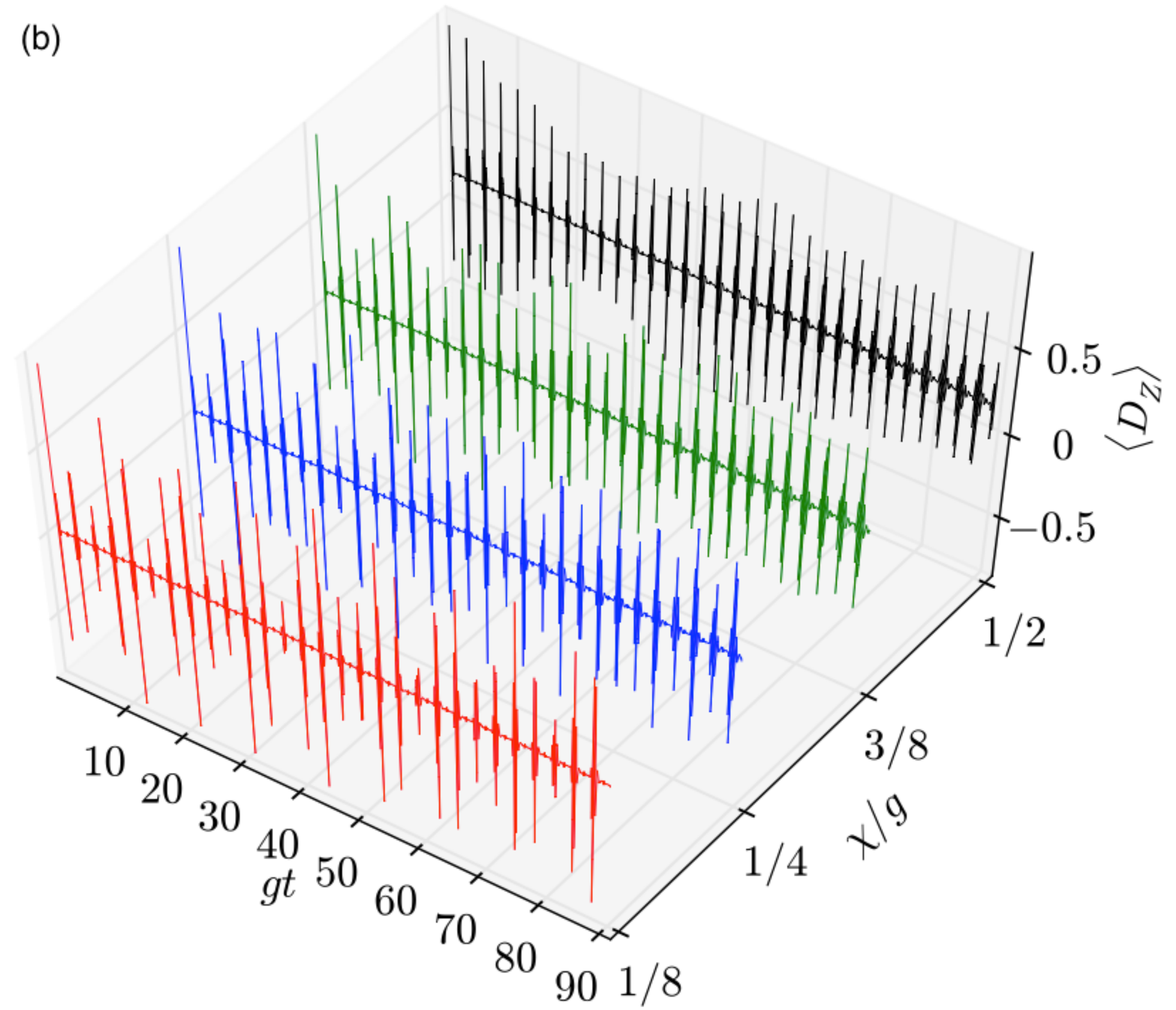} 
\caption{Plots of the time evolution of atomic inversion $\langle D_{Z} \rangle$ for an initially coherent state with $\langle \hat{n} \rangle=10$ photons. The parameters are $\delta=0$, $g=5\times 10^{-4}$, $\kappa-J=0$, with $h(n)=1+\frac{\chi}{\omega_{0}}n$, $f(n)=1$ (left panel) and  $f(n)=\sqrt{n}$ (right panel). The following cases are considered: $\chi/g= 1/8$, $1/4$, $3/8$, and $1/2$, corresponding to red, blue, green and black curves, respectively.}
\label{fig:atinversion1}
\end{center}
\end{figure}

The evolution of atomic excitation exhibits its characteristic collpase-and-revival behavior (explained in terms of the granular statistics of the field \cite{eberly}) but in a not-so-conventional fashion, depending on whether the nonlinear character of the medium is featured or whether the interplay of the atoms, together with the nonlinear coupling between them and the radiation field, is taken into account in our description. For instance, one can see from the sequence of graphs in Fig. \ref{fig:atinversion1} (a) that in the case of a linear coupling, for which $f(n)=1$, and under the condition $\kappa=J$, the appearance of revivals in the dynamics of atomic excitation tends to be more frequent insofar as the value of $\chi$ increases, and the profile of such revivals becomes sharpened and squeezed in the meanwhile. Following this sequence, the results also reveal the fact that the offset created by the constant term in the atomic excitation turns out to be highly dependent on $\chi$; the larger its value, the more significant its contribution to the deviation from the abscissa where $\langle D_{Z} \rangle =0$. On the other hand, adding a Buck-Sukumar-like function $f(n)=\sqrt{n}$ to the coupling between the atoms and the field fuels significantly the foregoing behavior as far as the profile and the number of revivals we observe, although in this case their frequency of appearance remains almost unchanged provided we keep ourselves confined to these restricted values of $\chi$ (see Fig. \ref{fig:atinversion1} (b)). Comparatively, the offset seems to be less sensitive to the combined effect of both nonlinear contributions.\\

Some structure in the overall atomic population profile can be found if the anharmonic regime of the cavity field, in combination with the interatomic interplay and the current intensity dependent atom-field coupling, is such that the condition $\chi/g \ll 1$ holds. Let us consider a situation in which the relationship between the aforesaid anharmonicity and the atomic interplay is such that $\chi =2(\kappa-J)$. This particular scenario allows us to simplify further our algebraic results and, more importantly,  quantify in some detail the dynamics of the atomic population when the above-mentioned influences are compounded. So, under the additional condition $\langle n \rangle \gg 1$ (i.e., in the limit of moderately high field intensities), one gets, in terms of what we have labeled, for the time being, as the scaled anharmonicity parameter $x=\chi/g$ (with $g$ being the coupling constant), the following effective approximations for the weighting amplitudes (\ref{eq:amplitudes}):
\begin{eqnarray}
\Lambda_{21}^{(x)} & \approx & 0, \\
\Lambda_{31}^{(x)} & \approx & \frac{1}{4(1+x^{2})(x^{2}+x\sqrt{1+x^{2}}+1)}, \\
\Lambda_{23}^{(x)} & \approx & \frac{1}{4(1+x^{2})(x^{2}-x\sqrt{1+x^{2}}+1)},
\end{eqnarray}
which are taken to be common to all oscillatory components in (\ref{eq:inversion}) around $\langle n \rangle$. Likewise, for the ones  associated with the offset's contribution we obtain
\begin{eqnarray}
\Lambda_{11}^{(x)} & \approx & -\frac{x^{2}+x\sqrt{1+x^{2}}}{4(x^{2}+x\sqrt{1+x^{2}}+1)^{3}}, \\
\Lambda_{22}^{(x)} & \approx & -\frac{x^{2}-x\sqrt{1+x^{2}}}{4(x^{2}-x\sqrt{1+x^{2}}+1)^{3}}, \\
\Lambda_{33}^{(x)} & \approx & 0.
\end{eqnarray}
We now proceed to provide an approximate version of the Rabi frequencies $\Omega_{31}^{(x,n)} $ and $\Omega_{23}^{(x,n)}$, regarded as functions of the independent variables $x$ and $n$, by expressing them in terms of their Taylor's series expansion about the point $(x,n)=(0,\langle n \rangle)$; this is so because, as functions of $n$, such frequencies are supposed to be strongly weighted around $\langle n \rangle$ by virtue of the Poissonian statistics of the initial coherent field. That is,
\begin{eqnarray}
\fl \Omega^{(x,n)}_{ij} & = & \Omega^{(0,\langle n \rangle )}_{ij}+\frac{\partial}{\partial n} \Omega^{(0,\langle n \rangle )}_{ij}(n-\langle n\rangle)+\frac{\partial}{\partial x} \Omega^{(0,\langle n \rangle )}_{ij}x \label{eq:series1} \\
\fl & & +\frac{1}{2!} \left [ \frac{\partial^{2}}{\partial n^{2}} \Omega^{(0,\langle n \rangle )}_{ij}(n-\langle n\rangle)^{2} +2\frac{\partial^{2}}{\partial n\partial x}\Omega^{(0,\langle n \rangle )}_{ij}(n-\langle n\rangle)x +\frac{\partial^{2}}{\partial x^{2}}\Omega^{(0,\langle n \rangle )}_{ij}x^{2} \right ]+O(x^{3},n^{3}). \nonumber 
\end{eqnarray}
Based on (\ref{eq:rabi2}) and (\ref{eq:rabi3}), together with the set of equations (\ref{eq:cardanosqt}), the corresponding expansion coefficients, up to second order in $x$ and $n$, take the form
\begin{eqnarray}
\fl \Omega^{(0,\langle n \rangle )}_{31} & = &  \Omega^{(0,\langle n \rangle )}_{23} \approx g\left (2\langle n \rangle + 3\right ),  \label{eq:domei1}\\ 
\fl \frac{\partial}{\partial n}\Omega_{31}^{(0,\langle n \rangle )} & = & \frac{\partial}{\partial n} \Omega^{(0,\langle n \rangle )}_{23} \approx 2g,\\
\fl \frac{\partial}{\partial x} \Omega^{(0,\langle n \rangle )}_{31} & = & -\frac{\partial}{\partial x} \Omega^{(0,\langle n \rangle )}_{23} \approx 3g,\\
\fl \frac{\partial^{2}}{\partial n\partial x}\Omega^{(0,\langle n \rangle )}_{31} & = & - \frac{\partial^{2}}{\partial n\partial x}\Omega^{(0,\langle n \rangle )}_{23} = \frac{3g(2\langle n \rangle^{2}+8\langle n \rangle +7)}{\Delta_{\langle n \rangle}^{+2}} \approx 0, \quad \textrm{for} \quad \langle n \rangle \gg 1, \\
\fl \frac{\partial^{2}}{\partial n^{2}} \Omega^{(0,\langle n \rangle )}_{31} & = &  \frac{\partial^{2}}{\partial n^{2}} \Omega^{(0,\langle n \rangle )}_{23} \approx \frac{\sqrt{2} g}{\Delta_{\langle n \rangle}^{+3/2}},  \\
\fl \frac{\partial^{2}}{\partial x^{2}}\Omega^{(0,\langle n \rangle )}_{31} & = & \frac{\partial^{2}}{\partial x^{2}}\Omega^{(0,\langle n \rangle )}_{23} \approx \frac{4g(\langle n \rangle +1)^{2}}{\sqrt{2 \Delta^{+}_{\langle n \rangle}}}. \label{eq:domef1}
\end{eqnarray}
Here, aside from the homogeneous terms $\Omega^{(0,\langle n \rangle )}_{ij}$, $\frac{\partial}{\partial x} \Omega^{(0,\langle n \rangle )}_{ij}$, and $\frac{\partial^{2}}{\partial x^{2}} \Omega^{(0,\langle n \rangle )}_{ij}$, each derivative of $\Omega^{(x,n)}_{ij}$ with respect to $n$ is known to define different time scales \cite{yoo,averbuch}: the first order derivatives are responsible for the revival behavior whose periodicity is easily determined to be $t_{R}=2\pi /|\frac{\partial}{\partial n}\Omega^{(0,\langle n \rangle)}_{ij}|$, whereas, at least in the regime we are working on, the second and higher order ones give rise to dephasing effects that become significant over much longer time scales. 

\begin{figure}[h!]
\begin{center}
\includegraphics[width=7.5cm, height=4.5cm]{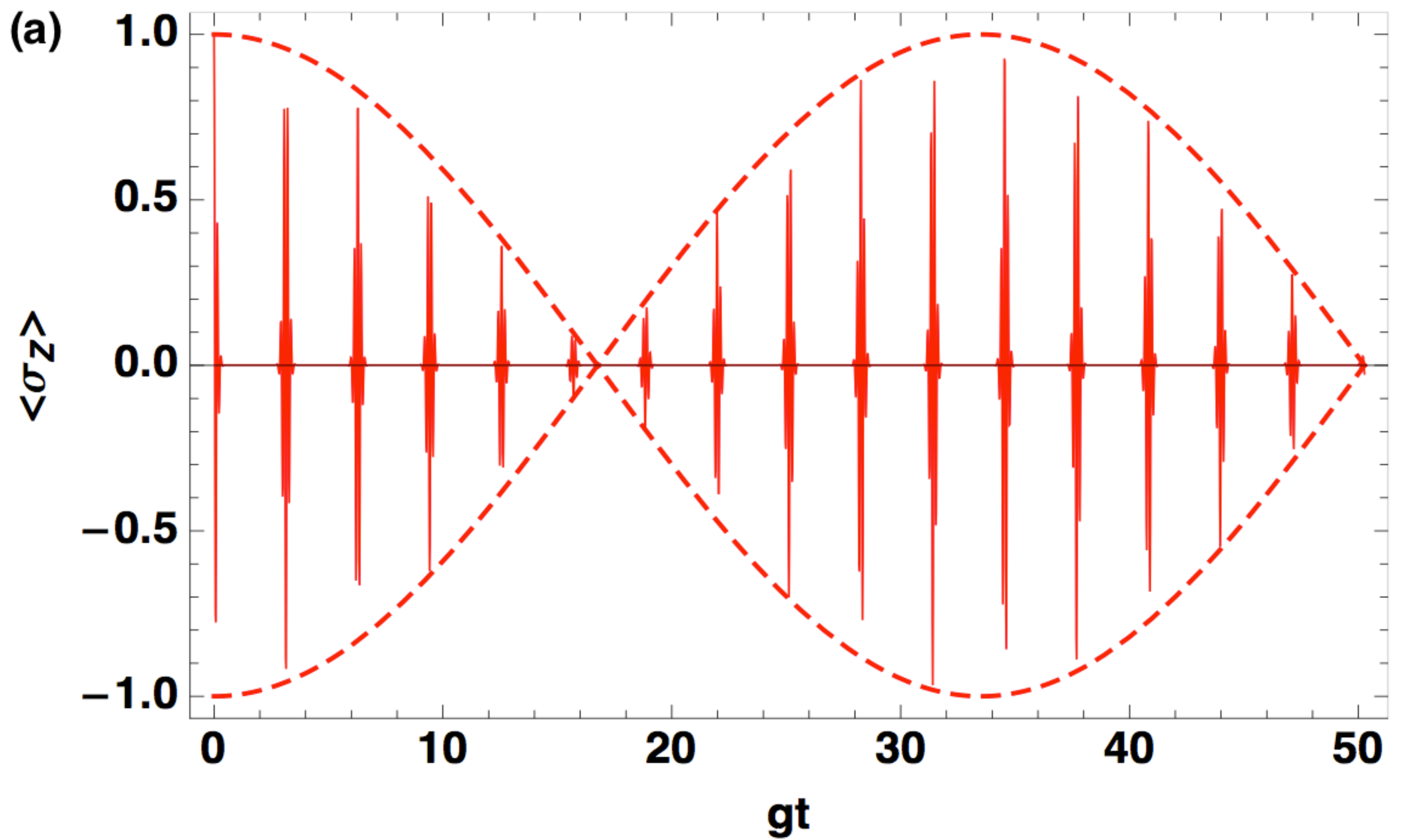} 
\includegraphics[width=7.5cm, height=4.5cm]{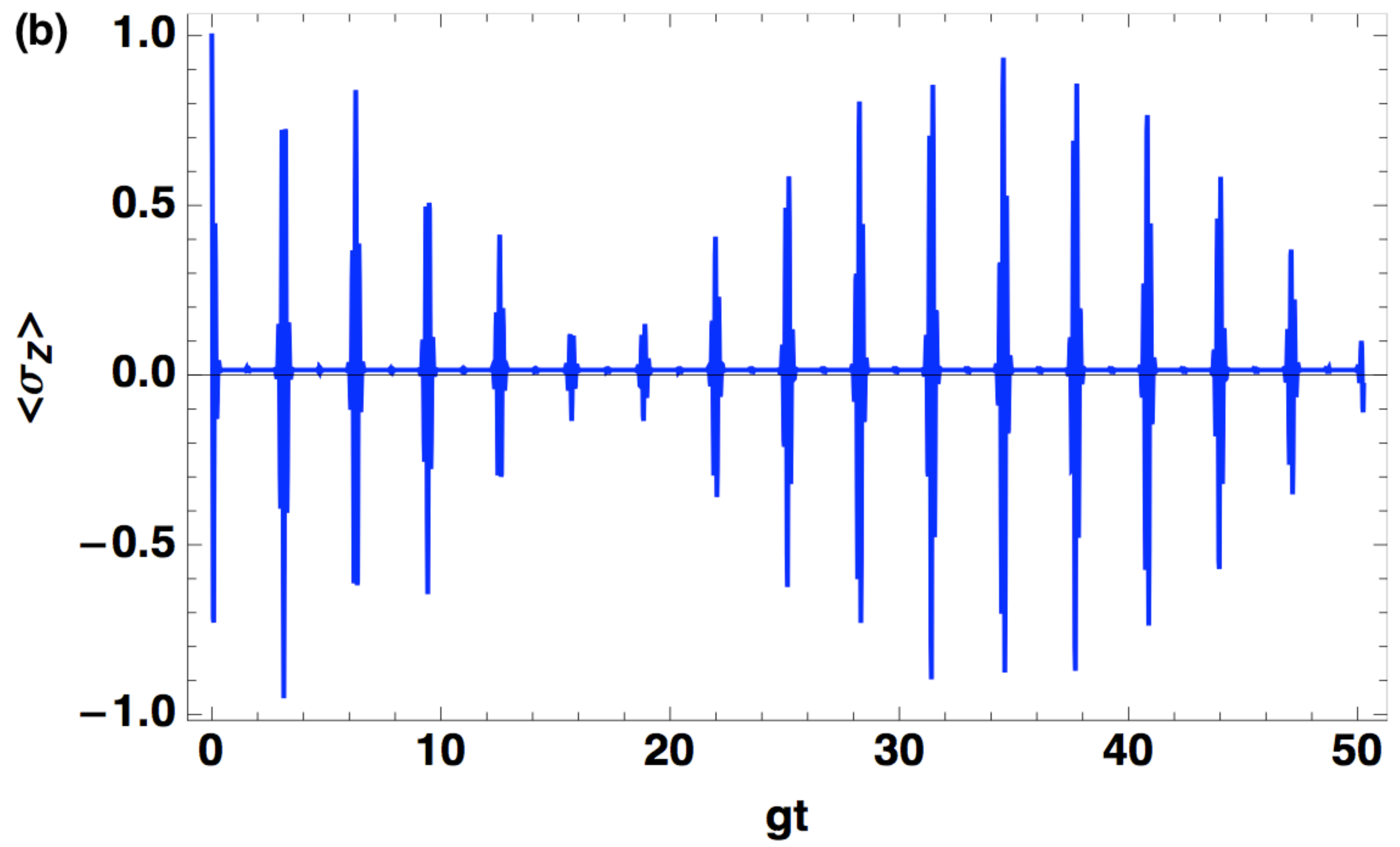} 
\caption{ Plots of atomic excitation $\langle D_{Z} \rangle$ calculated from Eqs. (\ref{eq:invapproxchi}) (red line) and (\ref{eq:inversion}) (blue line) in the Buck-Sukumar regime ($f(n)=\sqrt{n}$) and interacting atoms within a Kerr-like cavity such that $\chi/g=2(\kappa-J)/g=1/32$, with the field being initially in a coherent state having an  average photon number $\langle n \rangle =20$. }
\label{fig:atinversion1chi}
\end{center}
\end{figure}

So, after substituting (\ref{eq:domei1})-(\ref{eq:domef1}) into (\ref{eq:series1}), we arrive, up to first order in $n$, at the sought approximations
\begin{eqnarray}
\Omega_{31}^{(n)} & \approx & g\left (2n+3(1+x)+\varphi_{\langle n \rangle}x^{2} \right), \\
\Omega_{23}^{(n)} & \approx & g\left (2n+3(1-x)+\varphi_{\langle n \rangle}x^{2} \right ),
\end{eqnarray}
where $\varphi_{\langle n \rangle} \approx \sqrt{2} (\langle n \rangle+1)^{2}/\sqrt{\Delta_{\langle n \rangle}^{+}} $ is an average-photon-number dependent phase shift; this behaves approximately as a linearly increasing function of $\langle n \rangle$ for $f(n)=\sqrt{n}$, hence even second order contributions of $x$ cannot be ignored for sufficiently high field intensities. On the basis of this set of approximations, we finally work out that the atomic inversion takes the form 
\begin{eqnarray}
\fl \langle \Phi(t)|D_{Z}|\Phi (t) \rangle  & \approx & \Lambda_{11}^{(x)}+\Lambda_{22}^{(x)}+e^{-2\langle n \rangle \sin^{2}(\tau)} \bigg \{ \Lambda_{31}^{(x)}\cos \left[ 3(1+x)\tau +\varphi_{\langle n \rangle}x^{2}\tau+\langle n \rangle \sin(2\tau) \right]  \nonumber \\
& & + \Lambda_{23}^{(x)} \cos \left[ 3(1-x)\tau +\varphi_{\langle n \rangle}x^{2}\tau+\langle n \rangle \sin(2\tau) \right] \bigg \}, \label{eq:invapproxchi}
\end{eqnarray}
where $\tau =gt$. In Fig. \ref{fig:atinversion1chi} we show, within the scaled interval $0\le \tau \le 16 \pi$, the outcome of this approximation (frame (a)) that is in agreement with the exact one coming directly from Eq. (\ref{eq:inversion}) (frame (b))  whenever $ \tau \ll \sqrt{2}\Delta_{\langle n \rangle}^{+3/2}/\langle n \rangle \approx 4\langle n \rangle^{2}$ holds; needless to say, this is so since the removal of second order contributions of $n$ in the Taylor series demands such a restriction. The evident aspect of the atomic evolution is that it exhibits a kind of beating behavior enclosing, in turn, a series of revivals of shorter duration. From Eq. (\ref{eq:invapproxchi}), the periodicity of the beating modulation, as well as the collapse and revival times corresponding to the confined revivals, can be approximately quantified. In this regard, we get $\tau \approx m\pi/3x=gm\pi/3\chi$, with $m$ being an integer, for the modulation frequency, whereas $\tau_{R}\approx \pi$ and $\tau_{C} \approx 1/\sqrt{2\langle n \rangle}$ for the inner revival and collapse times, respectively. This set of calculations were performed by considering a scaled anharmonicity parameter  $x \approx 0.03$ and a   coherent field with $\langle n \rangle =20$ photons. \\

It is worth highlighting the fact that even a very slight and/or inconspicuous deviation from the standard cavity produced by the Kerr-type nonlinearity, necessarily combined with an intensity dependent atom-field coupling, can produce a somewhat well-structured population dynamics. On the the contrary, if the the nonlinear character of the atom-field coupling is switch off, i.e., $f(n)=1$, the overall behavior looks like that in Fig. \ref{fig:atinversion1} (a) and seems to be, in turn, qualitatively similar to the one-atom case reported in the work of G\'ora and Jedrzejek \cite{gora}, a regime under which the structures described above do not take place. And, as mentioned at the outset of this work, our model not only goes beyond the one-atom case to consider the one of two interacting atoms but also would give us the possibility of exploring any other regime and/or physical scenario in which more complex atom-field couplings may be engineered \cite{moya}.

\subsubsection{Case 2. Standard cavity: $h(\hat{n})=1$}

\begin{figure}[h!]
\begin{center}
\includegraphics[width=7.5cm, height=7cm]{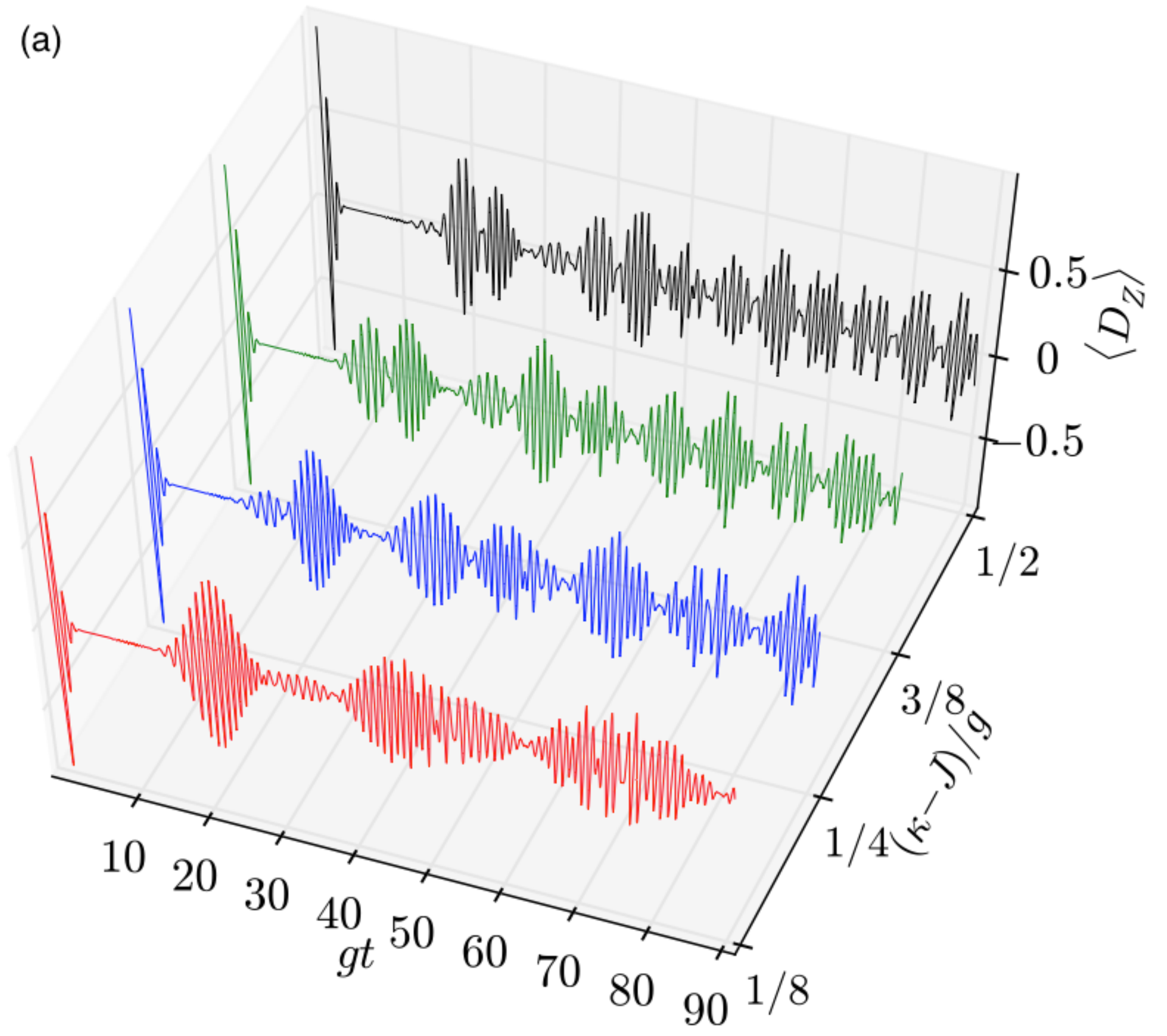} 
\includegraphics[width=7.5cm, height=7cm]{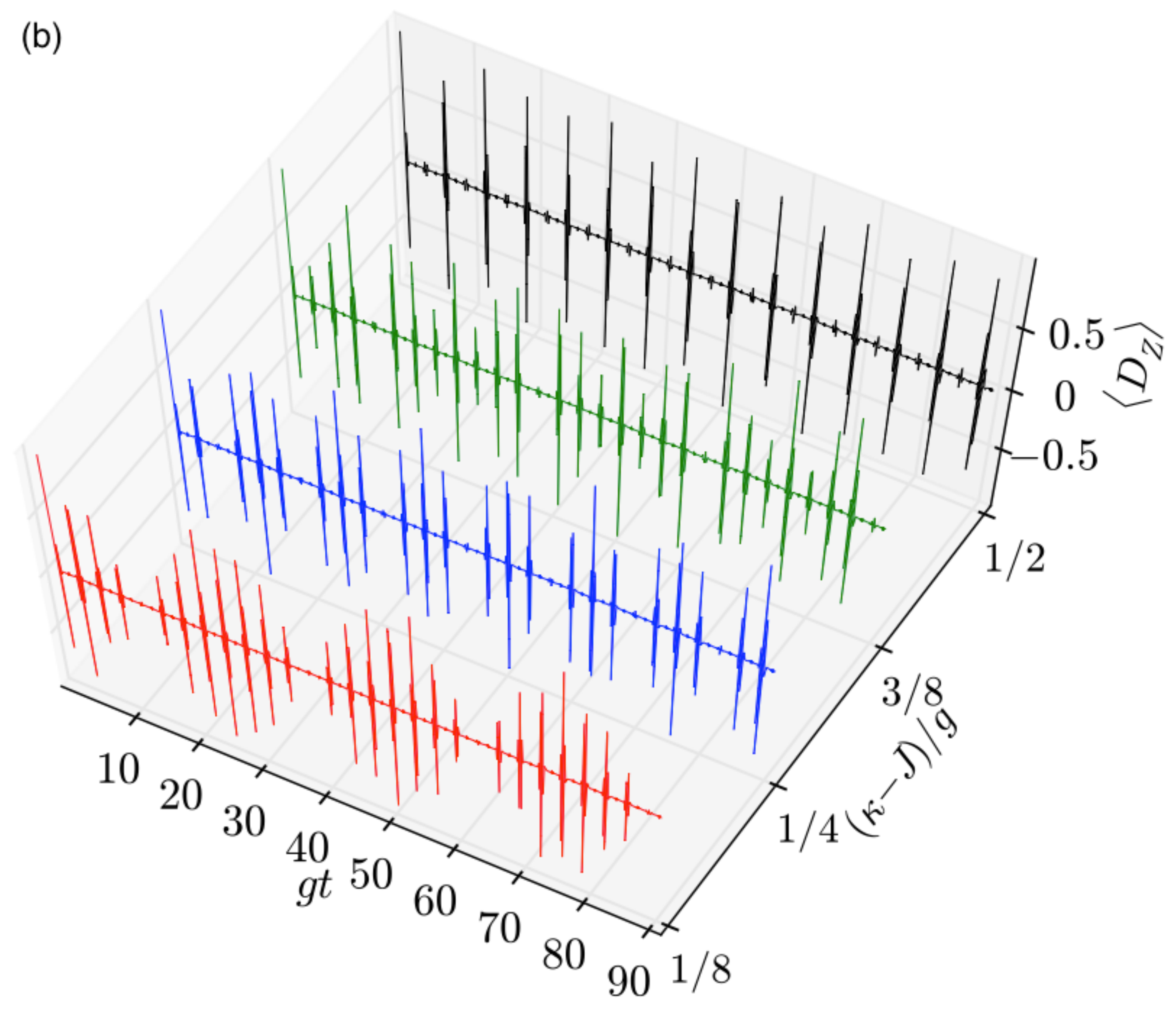} 
\caption{Plots of the time evolution of atomic inversion $\langle D_{Z} \rangle$ for an initially coherent state with $\langle \hat{n} \rangle=10$ photons. The parameters are $\delta=0$ and $\chi=0$, with $h(n)=1$, $f(n)=1$ (left column) and  $f(n)=\sqrt{n}$ (right column). The following cases are considered: $(\kappa -J)/g= 1/8$, $1/4$, $3/8$, and $1/2$, corresponding to red, blue, green, and black curves, repsectively.}
\label{fig:atinversion2}
\end{center}
\end{figure}

As to the standard cavity field, for which $\chi=0$, Cardano's formulae become
\begin{eqnarray}
Q_{n} & = &  -\frac{2}{9} \left( 2(\kappa-J)^{2}+3 g^{2}\Delta_{n}^{+} \right), \label{eq:qnkj} \\
\theta_{n} & = & \cos^{-1} \left( \frac{1}{\sqrt{2}} \frac{(\kappa-J) \left (4(\kappa-J)^{2}+9g^{2}\Delta_{n}^{+} \right)}{(2(\kappa-J)^{2}+3g^{2}\Delta_{n}^{+})^{3/2}} \right ). \label{eq:tetankj}
\end{eqnarray}
So, using these expressions enables us to deploy Eq. (\ref{eq:inversion}) in a way such that it can be viewed as a function of the difference $\kappa-J$, as we have mentioned before. The dynamics of the atomic occupation is displayed in Fig. \ref{fig:atinversion2} for $(\kappa-J)/g=1/8$, $1/4$, $3/8$ and $1/2$. In this regime, concerning the case in which $f(n)=1$ (see Fig. \ref{fig:atinversion2} (a)), the atomic system exhibits a somewhat similar collapse-and-revival behavior to that in the well-known standard two-atom Jaynes-Cummings model \cite{deng}, save for the fact that it is observed a faster blurring of the train of revivals with the increase of the difference $\kappa-J$. However, we see a completely different pattern when both the dipole-dipole-and-Ising interactions and the Buck-Sukumar coupling ($f(n)=\sqrt{n}$) are compounded (see Fig. \ref{fig:atinversion2} (b)). It is interesting to note that both contributions, working together, foster the formation of well-structured profiles of beats composed of a number of inner revivals, as seen, for instance, more remarkably from the red and blue graphs in Fig. \ref{fig:atinversion2} (b) corresponding to the cases when $(\kappa-J)/g=1/8$ and $1/4$, respectively; one sees that this is not an exclusive behavior of the Kerr effect observed previously. We also point out that neither the dipole-dipole-and-Ising interactions nor its combination with the nonlinear atom-field coupling play a proponderant role in the atomic excitation's offset.\\

\begin{figure}[h!]
\begin{center}
\includegraphics[width=7.5cm, height=4cm]{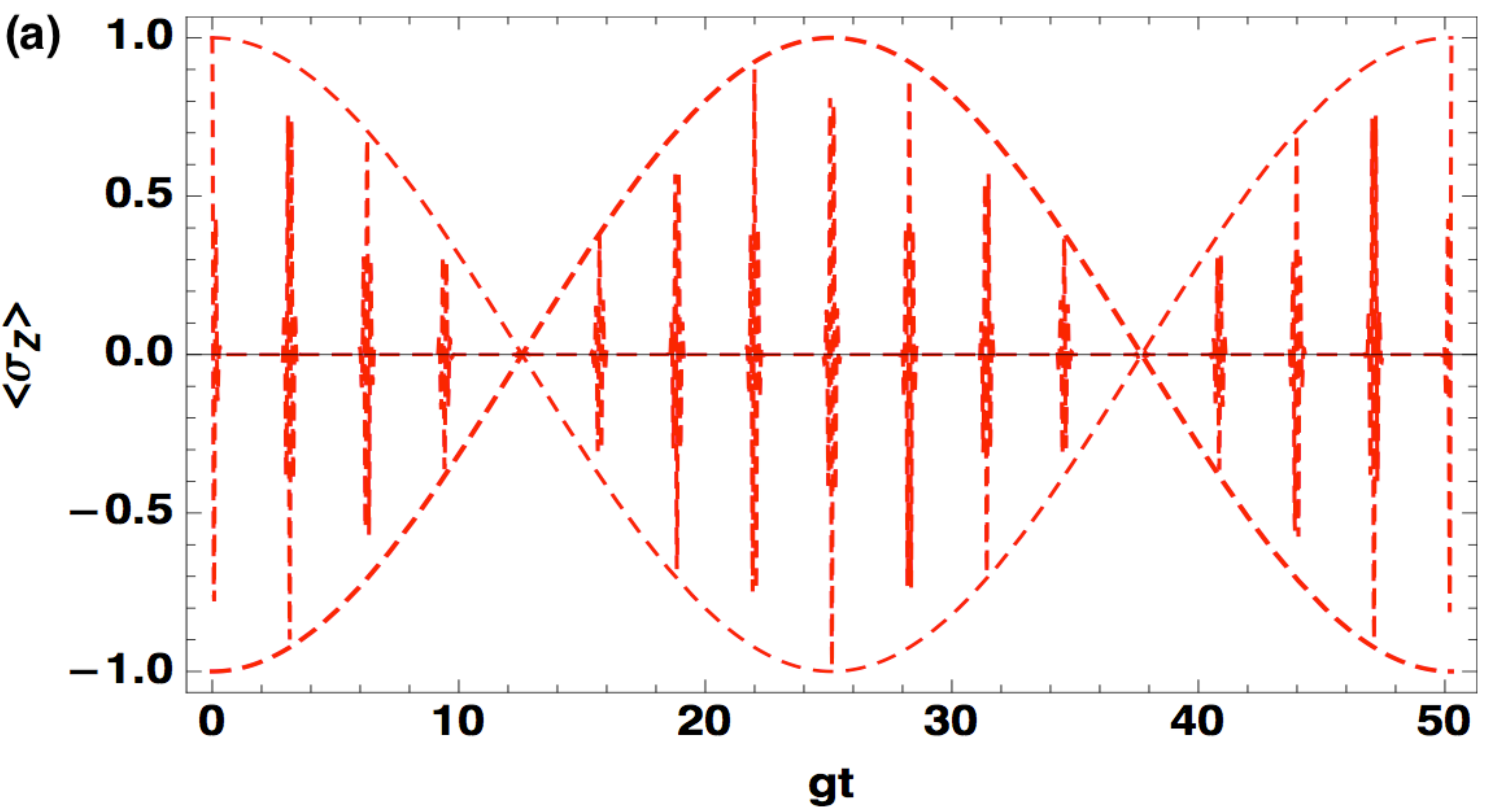} 
\includegraphics[width=7.5cm, height=4cm]{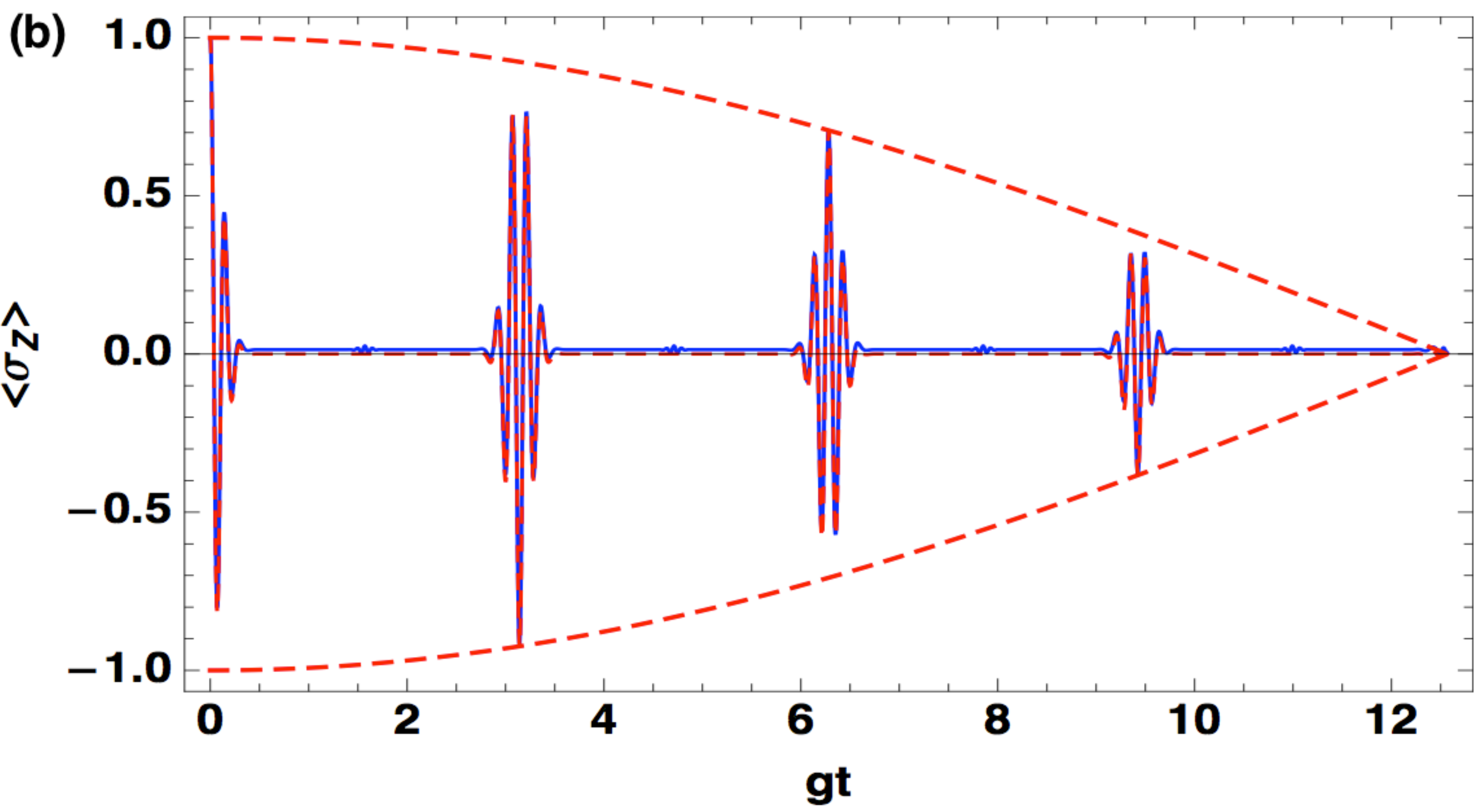} 
\includegraphics[width=7.5cm, height=4cm]{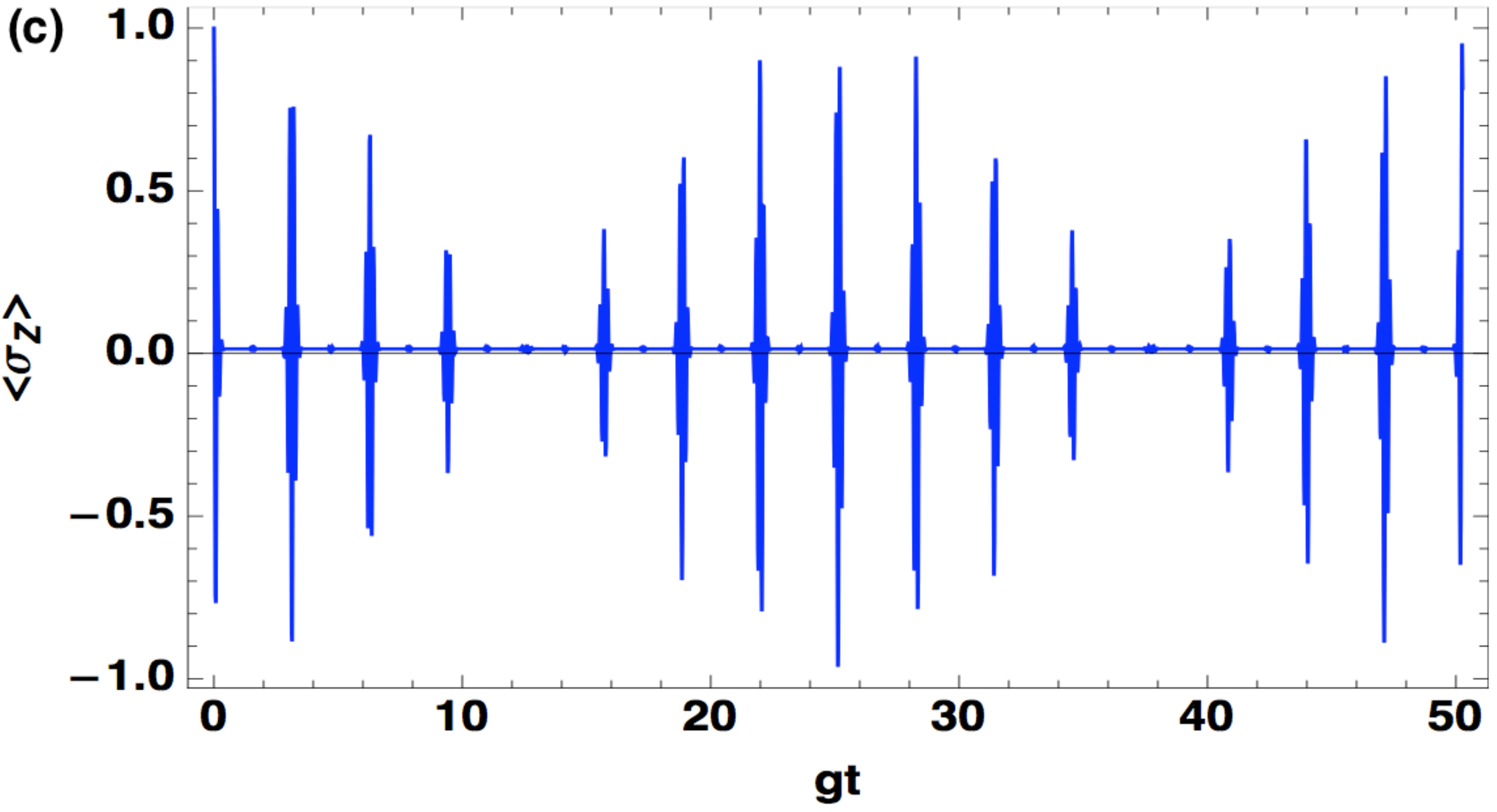} 
\includegraphics[width=7.5cm, height=4cm]{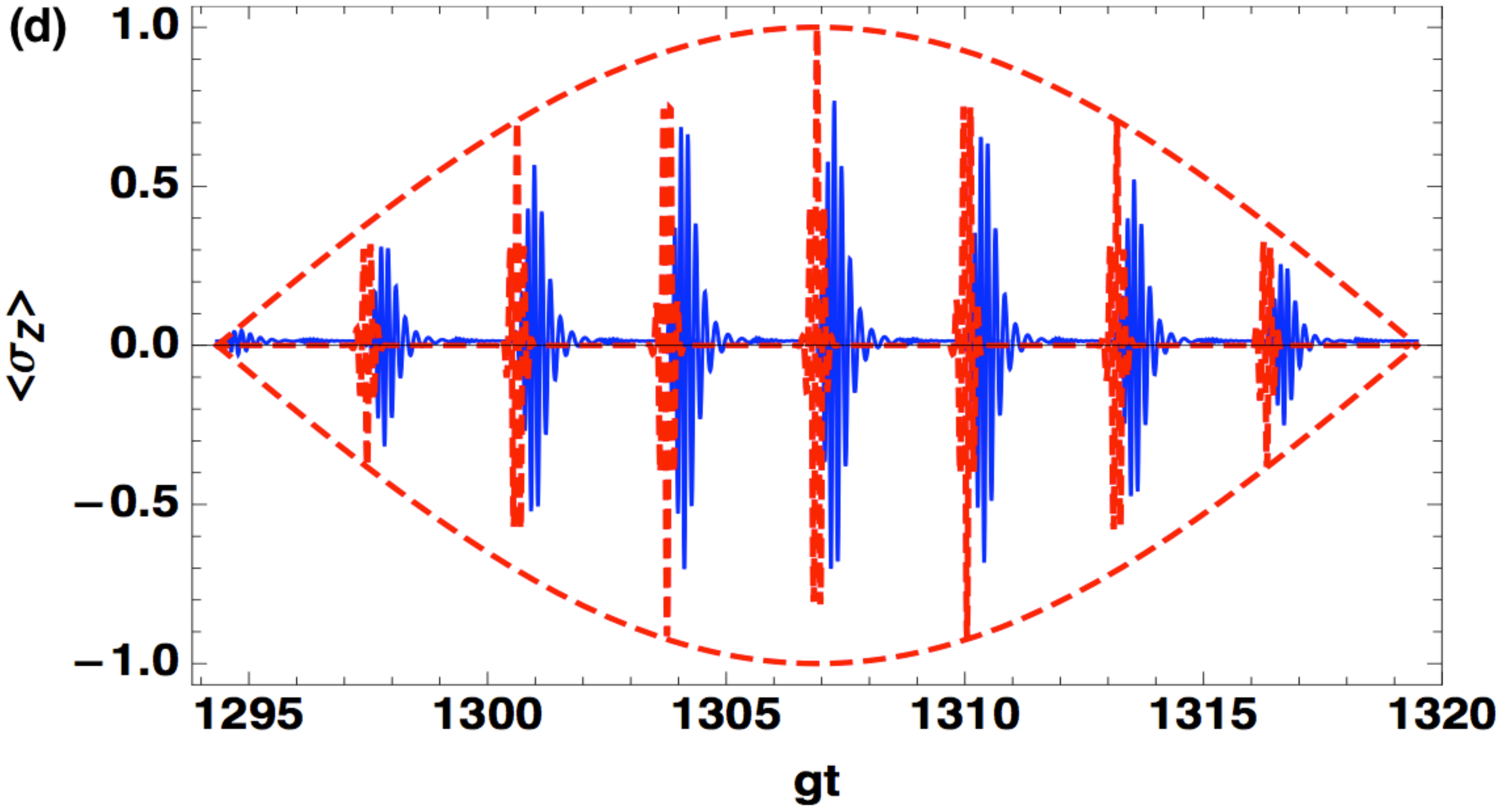} 
\caption{Plots of atomic excitation $\langle D_{Z} \rangle$ calculated from Eqs. (\ref{eq:inversion}) (blue, continuous line) and (\ref{eq:invapproxkj2}) (red, dashed line) in the Buck-Sukumar regime ($f(n)=\sqrt{n}$) and interacting atoms ($(\kappa-J)/g=1/8$) with the field being initially in a coherent state with average photon number $\langle n \rangle =20$. Frames (a) and (c) show the evolution of atomic excitation within the time interval $0\le gt \le16 \pi$. Frames (b) and (d) encompass two widely separated time intervals, $0\le gt \le 4\pi$ and $412\pi \le gt \le 420 \pi$, respectively, so as to emphasize the limitation of the approximation (\ref{eq:invapproxkj2}).}
\label{fig:atinversion3}
\end{center}
\end{figure}

Motivated by the above results involving both the nonlinear atom-field coupling ($f(n)=\sqrt{n}$) and the interatomic interplay within the regime $0<(\kappa-J)/g < 1$ (which is, in fact, slightly less restrictive than that of the Kerr-like cavity) where the evolution of atomic excitation reveals what  seems to be a regular and well-structured profile, we undertake the task of looking for a suitable approximation in order to gain some insight into such a behavior. It turns out that, for $\langle n \rangle \gg 1$, the weighting amplitudes given by (\ref{eq:amplitudes}) are found to be well approximated as follows: $\Lambda_{21}^{(n)}\approx 0$ and $\Lambda_{31}^{(n)} \approx \Lambda_{23}^{(n)}\approx 1/4$. On the basis of Eqs. (\ref{eq:qnkj}) and (\ref{eq:tetankj}), let us now make a Taylor expansion of the surviving Rabi frequencies (\ref{eq:rabi2}) and (\ref{eq:rabi3}) in $(\kappa-J)/g$ and $n$ by employing Eq. (\ref{eq:series1}) to get explicitly the corresponding expansion coefficients up to second order in $x$ and $n$:
\begin{eqnarray}
\Omega^{(0,\langle n \rangle )}_{31} & = &  \Omega^{(0,\langle n \rangle )}_{23} \approx g\left (2 \langle n \rangle + 3\right ), \label{eq:domei} \\ 
\frac{\partial}{\partial n}\Omega_{31}^{(0,\langle n \rangle )} & = & \frac{\partial}{\partial n} \Omega^{(0,\langle n \rangle )}_{23} \approx 2g,\\
\frac{\partial}{\partial x} \Omega^{(0,\langle n \rangle )}_{31} & = & -\frac{\partial}{\partial x} \Omega^{(0,\langle n \rangle )}_{23} \approx g,\\
\frac{\partial^{2}}{\partial n\partial x}\Omega^{(0,\langle n \rangle )}_{31} & = & \frac{\partial^{2}}{\partial n\partial x}\Omega^{(0,\langle n \rangle )}_{23} \approx 0,\\
\frac{\partial^{2}}{\partial n^{2}} \Omega^{(0,\langle n \rangle )}_{31} & = &  \frac{\partial^{2}}{\partial n^{2}} \Omega^{(0,\langle n \rangle )}_{23} \approx \frac{g}{2(\langle n \rangle+3/2)^{3}}, \label{eq:tscalen} \\
\frac{\partial^{2}}{\partial x^{2}}\Omega^{(0,\langle n \rangle )}_{31} & = & \frac{\partial^{2}}{\partial x^{2}}\Omega^{(0,\langle n \rangle )}_{23} \approx \frac{g}{2\langle n\rangle +3}. \label{eq:domef}
\end{eqnarray}
Here, we have set $x=(\kappa-J)/g$ and all derivatives are evaluated at the point $(0,\langle n \rangle)$; additionally, we have applied the fitting approximation $\Delta_{\langle n \rangle}^{+}\approx \frac{1}{2}(2 \langle n \rangle +3)^{2}$. At this point, as opposed to the Kerr-cavity case, given the inverse proportionality of (\ref{eq:domef}) as a function of $\langle n \rangle $ it makes sense to retain only contributions up to first order in $x$ in the expansion, thereby getting 
\begin{eqnarray}
\Omega_{31}^{(n)} & \approx & g\left( 2n+3+\frac{\kappa-J}{g} \right),\\
\Omega_{23}^{(n)} & \approx & g\left( 2n+3-\frac{\kappa-J}{g} \right),
\end{eqnarray}
where we have also found it convenient to cut the series up to first order terms in $n$, as long as we restrict ourselves to proper time scales. On substituting all the foregoing approximate expressions into Eq. (\ref{eq:inversion}), we obtain the sought result
\begin{equation}
\langle \Phi(t)|D_{Z}|\Phi (t) \rangle  \approx \cos \left ((\kappa-J)\tau/g \right) \sum_{n=0}^{\infty}P_{n} \cos \left((2n+3)\tau \right),
\label{eq:invapproxkj}
\end{equation}
where we have set $\tau=gt$. This expression is, in turn, summed exactly to arrive at the following closed form
\begin{equation}
\langle \Phi(t)|D_{Z}|\Phi (t) \rangle  \approx e^{-2\langle n\rangle \sin^{2}(\tau)}\cos \left ((\kappa-J)\tau /g \right) \cos \left[3\tau+\langle n \rangle \sin(2\tau) \right].
\label{eq:invapproxkj2}
\end{equation}
Based on this approximate result, the evolution of atomic excitation can be understood in terms of the product of what we can identify as an envelope factor, $e^{-2\langle n\rangle \sin^{2}(\tau)} \cos \left ((\kappa-J)\tau /g \right)$, and a more rapidly oscillating cosine function. The essential feature of the former is twofold as depicted in Fig. \ref{fig:atinversion3} (a). I.e., on the one hand, it transpires that the exponential part of it  takes part in enveloping the inner revivals generated by the term of higher frequency $\cos \left[3\tau+\langle n \rangle \sin(2\tau) \right]$; incidentally, the argument of the latter, in contrast to the one-atom model, contains an additional time dependent term, $3\tau$, that is in fact responsible for the alternating phase inversion of successive revivals at the first stages of atomic evolution, as seen in Fig. \ref{fig:atinversion3} (b) where, for the sake of comparison, we have also depicted the exact result obtained by means of Eq. (\ref{eq:inversion}) for the particular case of $(\kappa -J)/g=1/8$ in the scaled interval $0\le gt \le 4\pi$. On the other hand, it follows that the dipole-dipole and/or Ising interactions constrain the overall modulation amplitude (outer, red, dashed line in Fig. \ref{fig:atinversion3} (a)) via the enveloping term of lower frequency, $\cos \left ((\kappa-J)\tau /g \right)$, thereby giving rise to the above-mentioned characteristic train of beats; their periodicity is easily determined to be held at times $\tau = gm\pi/(\kappa-J)$, with $m$ being an integer. We have plotted in Fig. \ref{fig:atinversion3} (c) the exact result (blue, continuous line) in the same interval as that of the approximate version of it in Fig. \ref{fig:atinversion3} (a), i.e., $0\le gt \le 16\pi$, where the qualitative behavior of the former is well reproduced by the latter even at such time scales. Finally, both the exact (blue, continuous line) and approximate (red, dashed line) results are displayed together in Fig. \ref{fig:atinversion3} (d) in a much more faraway time interval, where it is clear that the outcome of Eq. (\ref{eq:invapproxkj2}) (red, dashed line) fails at these far longer times. This noticeable discrepancy between the approximate and the exact results is due in part to the asymptotic approximations involving the weighting amplitudes and the number-dependent frequency shifts  $\Delta_{n}^{+}$ given above and primarily to the removal of the second order contributions of $n$ in the Taylor expansion of the Rabi frequencies. Based on (\ref{eq:series1}) and (\ref{eq:tscalen}), as long as we restrict ourselves to time scales such that $\tau \ll \frac{2}{\langle n \rangle}\left(\frac{\partial^{2}\Omega_{ij}^{(0,\langle n\rangle)}}{\partial n^{2}} \right)^{-1} \sim  4 \langle n\rangle^{2}$, with $\tau=gt$, Eq. (\ref{eq:invapproxkj2}) gives us an illustrative and quite acceptable first approximation and becomes more and more accurate for longer periods of time as the coherent-state average photon number $\langle n\rangle$ increases. And, in exactly the same way as in the Kerr cavity case, one can extract from it some useful information concerning the inner collapse and revival times, namely, $\tau_{C} \approx 1/\sqrt{2 \langle n \rangle}$ and $\tau_{R} = \pi$, respectively, and, it is of some importance to remark that both of them turn out to halve their corresponding value in the one-atom case \cite{sukumar}. This specific set of calculations was carried out by considering the initial state of the field as being in a coherent state with $\langle n\rangle =20$ photons. \\

\subsection{Purity and concurrence}

From the point of view of quantum information, another feature that deserves to be examined is the  entanglement dynamics of the two interacting atoms. To do this, let us now suppose that our atomic system is initially prepared in the symmetric entangled state given, for a definite number of photons $n$, by Eq. (\ref{eq:state2}). In doing so, and following the same procedure as that outlined above for the atomic excitation, one finds that the state of the composite system evolves as follows:
\begin{equation}
|\Phi (t) \rangle = \sum_{n=0}^{\infty}\sum_{k=1}^{3} A_{n} \mathcal{D}_{k}^{(n)} (t)|\phi_{k}^{(n)} \rangle,
\end{equation}
where, again, the state of the cavity field is supposed to be in a general superposition of number states, each of them weighted by the set of amplitudes $A_{n}$, and the time-dependent coefficient takes now the form
\begin{equation}
\mathcal{D}_{k}^{(n)}(t) = \sum_{j=0}^{3} C_{j2}^{(n)} C_{jk}^{(n)} e^{-iE_{j}^{(n)}t}.
\label{eq:coefficient2}
\end{equation}
Thus, starting from the corresponding density operator of the atom-field system, namely,
\begin{equation}
\hat{\rho}_{AF} = |\Phi (t) \rangle \langle \Phi(t)| = \sum_{n,m}\sum_{j,k} A_{n}A_{m}^{\ast} \mathcal{D}_{k}^{(n)}(t)\mathcal{D}_{j}^{(m)\ast}(t) |\phi_{k}^{(n)} \rangle \langle \phi_{j}^{(m)}|,
\label{eq:densityaf}
\end{equation}
we first proceed to calculate the purity of the atomic system alone by means of the standard definition of this property
\begin{equation}
P(\hat{\rho}_{A}) = \Tr \{ \hat{\rho}^{2}_{A} \},
\label{eq:purityatom}
\end{equation}
where $\rho_{A}$ represents the reduced density operator of the atoms computed via 
\begin{equation}
\hat{\rho}_{A} = \Tr_{F} \{ \hat{\rho}_{AF} \},
\end{equation}
with $\Tr _{F}$ denoting the trace over the field variables. Hence, one obtains for the reduced density operator 
\begin{equation}
\hat{\rho}_{A} = \sum_{n}^{\infty} \sum_{j,k=1}^{3} A_{n+j-k}A_{n}^{\ast}\mathcal{D}_{k}^{(n+j-k)}(t)\mathcal{D}_{j}^{(n)\ast}(t) |\phi_{k} \rangle \langle \phi_{j}|,
\label{eq:rhoatoms}
\end{equation}
where the set of atomic states $\{ |\phi_{k} \rangle \}$ comes from having made use of a shorthand notation for the basis (\ref{eq:state1})-(\ref{eq:state3}), namely, we have labeled $|\phi_{k}^{(n)}\rangle =|\phi_{k}\rangle \otimes |n+k-1\rangle$ with the following correspondences: $|\phi_{1}\rangle =|e,e\rangle$, $|\phi_{2} \rangle =\left (|e,g\rangle +|g,e\rangle \right)/\sqrt{2}$, and $|\phi_{3} \rangle = |g,g\rangle$. Then, on substituting the last expression into (\ref{eq:purityatom}), we get
\begin{equation}
P(\hat{\rho}_{A}) = \sum_{j,k=1}^{3} \left |\sum_{n}^{\infty} A_{n+j-k}A_{n}^{\ast}\mathcal{D}_{k}^{(n+j-k)}(t)\mathcal{D}_{j}^{(n)\ast}(t) \right |^{2}.
\label{eq:purity-atoms}
\end{equation}
In this context, the expression for the purity given above can be understood as a tool giving an account of the degree of entanglement between the atomic system and the cavity field, with the latter being viewed as the environment, and of particular interest to us will be, again, a radiation field in a coherent superposition of number states.\\

\begin{figure}[h!]
\begin{center}
\includegraphics[width=7.6cm, height=5cm]{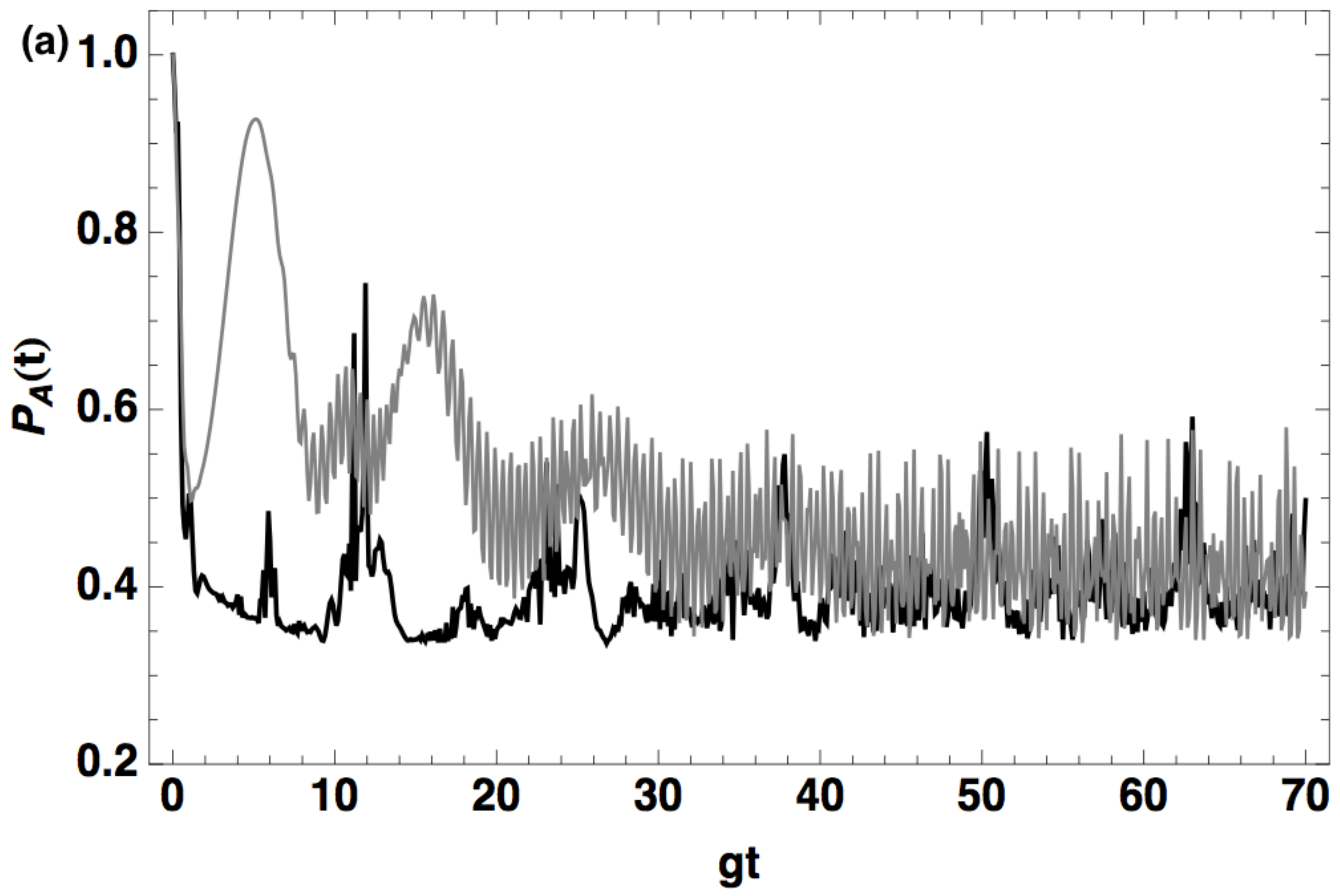} 
\includegraphics[width=7.6cm, height=5cm]{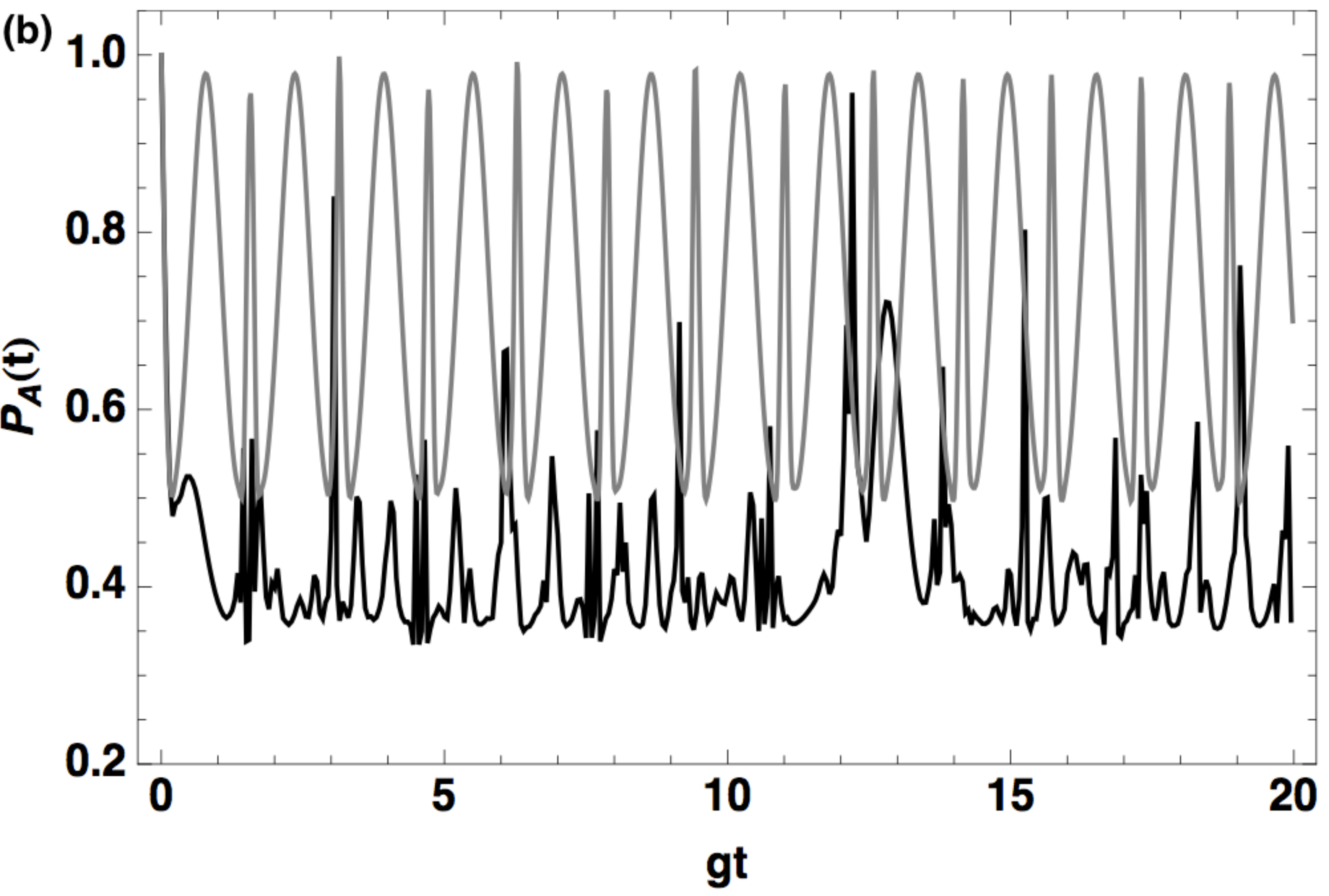} 
\caption{Plots of the evolution of atomic purity calculated from Eq. (\ref{eq:purity-atoms}) for the field being initially in a coherent state with an average photon number $\langle \hat{n} \rangle=10$. The parameters are $(\kappa-J)/g=1/4$ (gray line) and $\chi /g=1/4$ (black line), with $f(n)=1$ (Frame (a)) and $f(n)=\sqrt{n}$ (frame (b)). }
\label{fig:purity-cs}
\end{center}
\end{figure}

The evolution of the purity of the atomic system for a field initially prepared in a coherent state with $|\alpha|^{2}=\langle n \rangle =10$, is depicted in Fig. \ref{fig:purity-cs} encompassing the nonlinear regimes considered so far. More precisely, the figure aims to stress the role of the nonlinear atom-field coupling (i.e., the Buck-Sukumar regime in which the coupling function $f(n)=\sqrt{n}$) as well as the inherent nonlinearity of the Kerr-like cavity in the evolution of the aforementioned quantity. Figures \ref{fig:purity-cs} (a) and (b) show, respectively, the corresponding outcomes when either a linear or a nonlinear atom-field coupling is taken into consideration. Frame (a) indicates, for instance, that within the linear regime ($f(n)=1$) of the standard cavity the central system of interacting atoms ($(\kappa-J)/g=1/4$; gray curve) tends to be more robust against decoherence at the first stages of its evolution in opposition to the Kerr-like cavity case ($\chi/g=1/4$; black curve) with non-interacting atoms ($\kappa-J=0$). Even though both cases display a kind of spiky behavior as time elapses, in the Kerr-like cavity the appearance of small, distinguishable and almost regular spikes revelas a very weak attempt to restore the initial coherence of the atomic system. This behavior seems to be reinforced and much more frequent when the intensity dependent coupling is featured ($f(n)=\sqrt{n}$), as seen in Frame \ref{fig:purity-cs} (b) on a shorter time scale. I.e., we see that the atom-atom interaction in conjunction with the Buck-Sukumar coupling (gray curve) strengthens the possibility of recovering almost completely the initial phase information of the two atoms at certain and more regular time instants. When both the Kerr-like cavity and the nonlinear coupling are working together (frame (b); black curve), the atomic system becomes most of the time less robust against decoherence, but it exhibits, in a haphazard fashion, more frequent and pronounced spikes than those appearing in the case of linear coupling (see frame (a); black curve).

\begin{figure}[h!]
\begin{center}
\includegraphics[width=7.6cm, height=5cm]{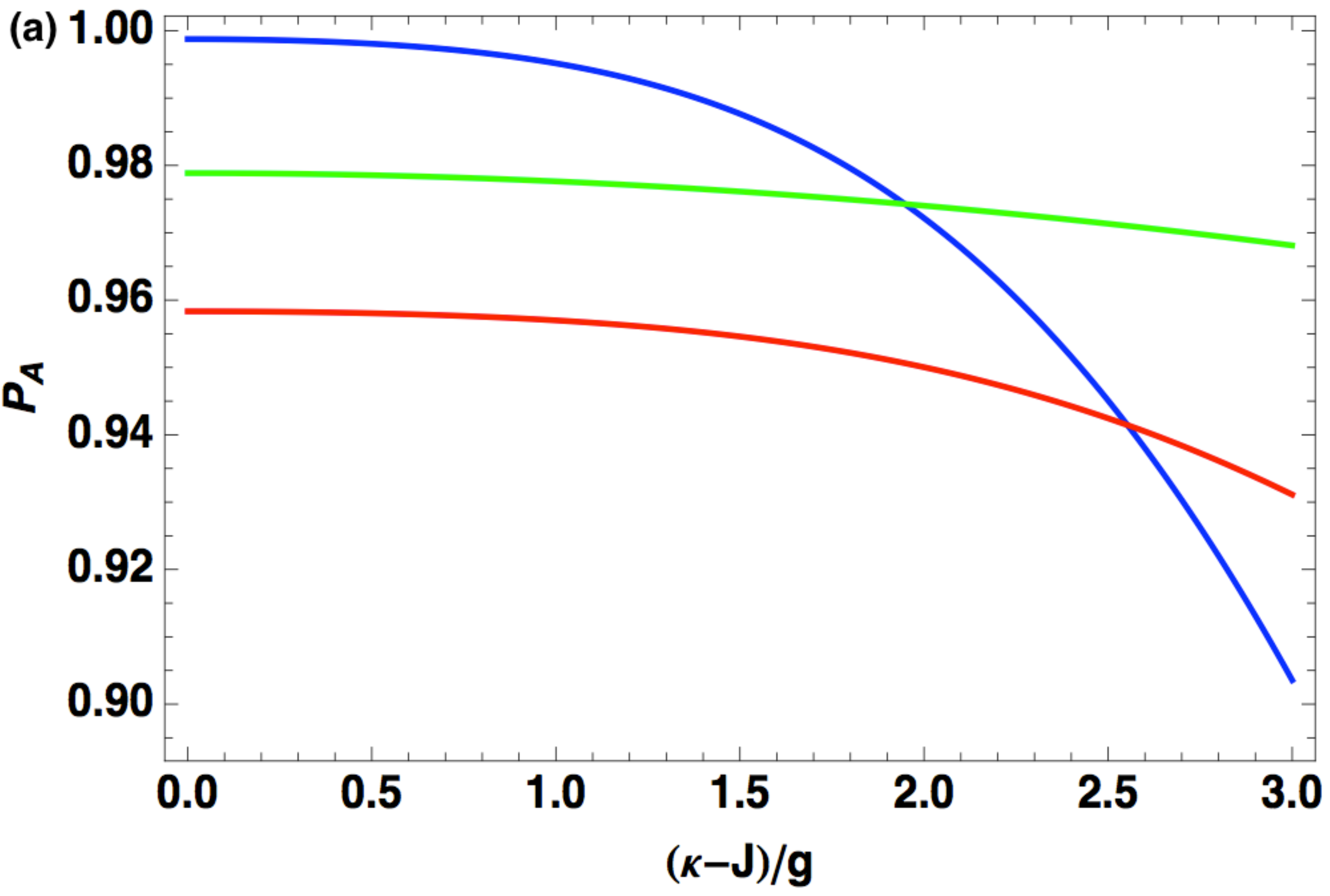} 
\includegraphics[width=7.6cm, height=5cm]{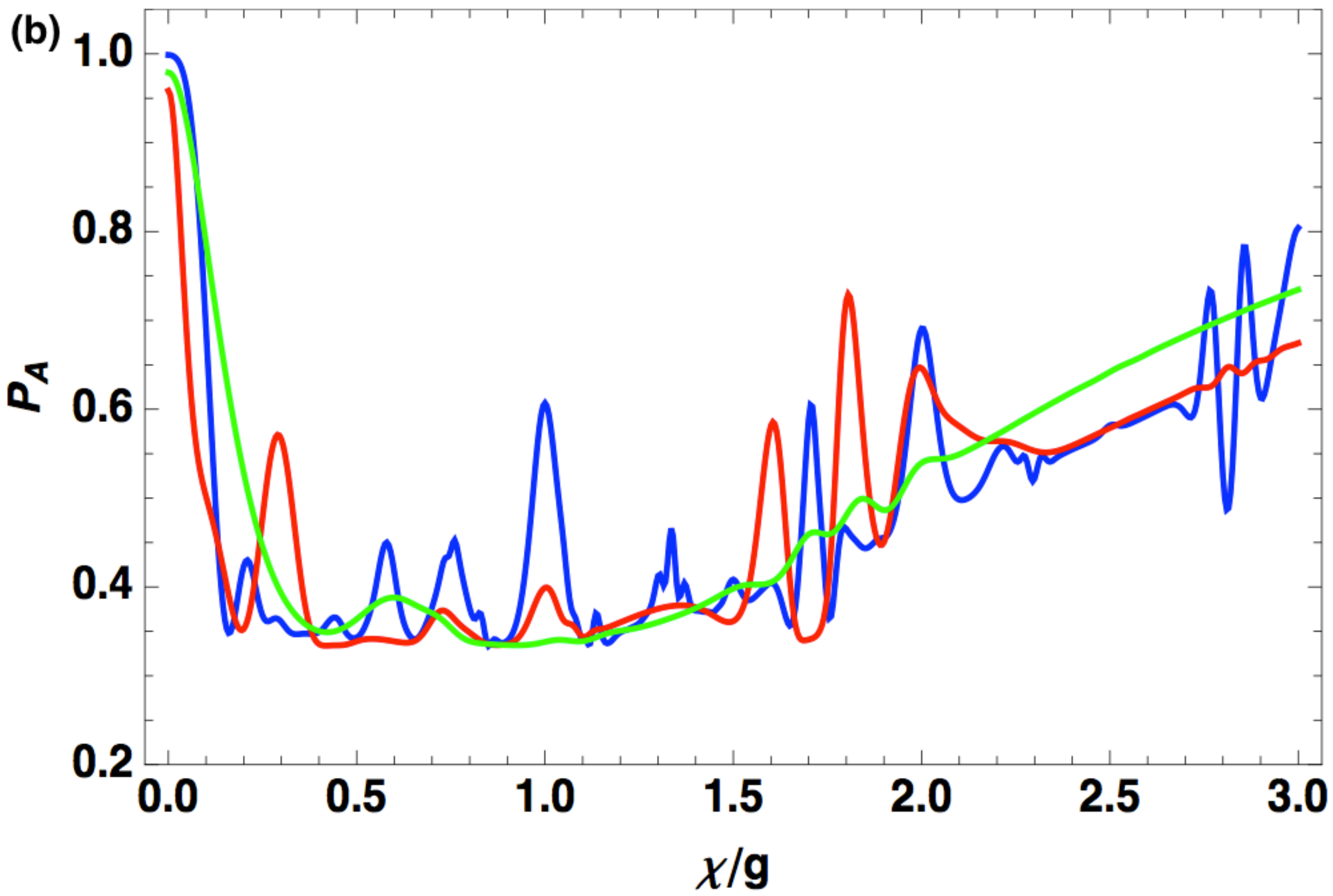} 
\caption{Atomic purity calculated from Eq. (\ref{eq:purity-atoms}) plotted, separately, as a function of $(\kappa-J)/g$ (frame (a)) and $\chi/g$ (frame (b)) for the time instants $gt=\pi/4$ (green), $\pi/2$ (red), and $\pi$ (blue). The field is initially in a coherent state with an average photon number $\langle \hat{n} \rangle=10$, and the interplay between it and the atoms takes place in the Buck-Sukumar regime ($f(n)=\sqrt{n}$). }
\label{fig:purity-cs2}
\end{center}
\end{figure}

Regarding the dipole-dipole and Ising-like atomic interactions, it has been verified that their influence upon the purity dynamics of the atoms themselves do not give rise to significant changes in the general behavior of it at least within the range $0<(\kappa-J)/g \le 1.5$, as shown in Fig. \ref{fig:purity-cs2} (a) for three different time instants, namely, $gt=\pi/4$, $\pi/2$ and $\pi$. Each of these correspond respectively, and approximately, to the second, third and fifth maximum in the evolution of purity shown in Fig. \ref{fig:purity-cs} (b) for the Buck-Sukumar-like coupling case (gray curve). So, as a function of the effective atomic parameter $(\kappa-J)/g$, we can see that the purity decreases gradually when we extend the range of values of $\kappa-J$; secondarily, in the regime when $(\kappa-J)/g>1.5$, the profile of the atomic purity as a function of time was also found to undergo an inconspicuous phase shift with respect to the case when $\kappa-J=0$ just at the first stages of the evolution. 

A somewhat different behavior is observed in the case of a Kerr-like cavity illustrated in Fig. \ref{fig:purity-cs2} (b) for the same time instants. From the figure we see that what started off being an almost pure atomic state ended up being a state that loses its initial coherence with the increase of $\chi/g$, and the decay of this atomic coherence turns out to be far more pronounced than that of the previous case. By contrast, far beyond the  range of values of the anharmonicity parameter considered  so far, say, when $\chi/g > 1.5$, the atomic purity seems to display a raising tendency as the aforesaid parameter increases. We surmise that this last conduct may be closely linked to the atomic population trapping effect \cite{buzek} induced by the presence of the nonlinear medium through a kind of detuning effect at such high values of $\chi/g$. \\

\begin{figure}[h!]
\begin{center}
\includegraphics[width=9cm, height=5.5cm]{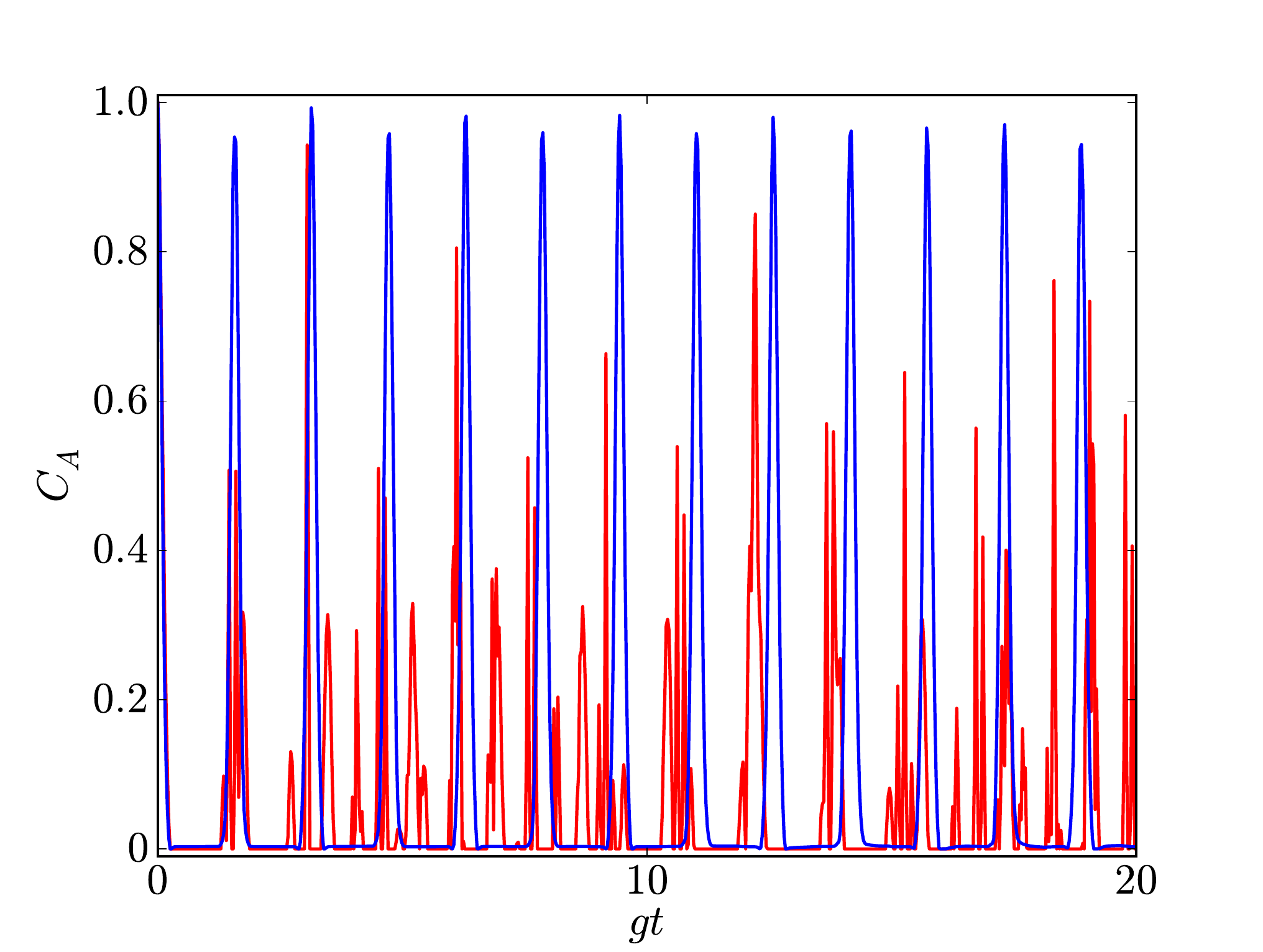} 
\caption{Plots of the evolution of atomic concurrence for the field being initially in a coherent state with an average photon number $\langle \hat{n} \rangle=10$. The parameters are $(\kappa-J)/g=1/4$ (blue line) and $\chi /g=1/4$ (red line), with $f(n)=\sqrt{n}$. }
\label{fig:concurrence}
\end{center}
\end{figure}
In order to close and complement the present subsection, let us report briefly the influence of the nonlinearities of the model on the degree of entanglement between the two atoms. This quantity can be suitably measured by dint of concurrence \cite{wootters}, which, for an arbitrary two-qubit system, is given by the expression $C(\rho_{A})=\max \{0, \lambda_{1}-\lambda_{2}-\lambda_{3}-\lambda_{4}\}$, where the $\lambda_{i}$'s are the square roots of the eigenvalues of the matrix $\rho_{A}\sigma_{y}^{(1)}\sigma_{y}^{(2)}\rho_{A}^{\ast}\sigma_{y}^{(1)}\sigma_{y}^{(2)}$ in a non-increasing order, with $\rho_{A}$ being the reduced density matrix of the atomic system. This calculation is entirely carried out numerically by solving the problem in the computational basis $\{ |g,g\rangle , |g,e\rangle, |e,g\rangle, |e,e\rangle \}$. 

Figure \ref{fig:concurrence} displays the concurrence dynamics of the atoms for the same set of parameters as in Fig. \ref{fig:purity-cs} (b) for purity. From the figure, one can see that the Buck-Sukumar coupling plays a preponderant role in bolstering the entanglement at certain times (see blue curve) that seem to be held approximately at integer multiples of $gt=\pi/2$ within the shown interval.
This is in agreement with some results encountered for the evolution of concurrence using a non-interacting qubit as a {\it probe}~\cite{gonzalez16} and it is a consequence of the periodicity of the dynamics generated by the Buck-Sukumar interaction.
A peculiar aspect that deserves to be noted is the fact that, on time domain, the appearance of the maxima of entanglement partially coincides with the one observed in purity dynamics just at such time instants; this is so inasmuch as the latter was found to reach its maximum values at integer multiples of $gt=\pi/4$, approximately, at least within the time interval pointed out in Fig. \ref{fig:purity-cs} (b). The presence of a Kerr-like medium, on the other hand, tends to undermine the entanglement between the atoms (red curve). This conduct is also somewhat similar to the one observed for purity and becomes more and more unpredictable as time elapses and with the increase of the ratio of the anharmonicity parameter to the coupling constant $\chi/g$. Once again, the dipole-dipole and Ising atomic interactions did not give rise to significant changes in  the foregoing evolution in the range of values $0< (\kappa-J)/g \le 1/2$ and under the present nonlinear atom-field coupling scheme.

\section{Evolution of the field: entropy and phase space}
In this last section we will try to describe the influence of the nonlinearities of the model upon the evolution of the cavity field by analyzing it in terms of its entropy and representation on phase space. Let us first examine the degree of mixture of the state of the field in terms of the von Neumann entropy  \cite{nielsen}
\begin{equation}
S = -\Tr \{\hat{\rho} \ln \hat{\rho} \},
\end{equation}
where $\hat{\rho}$ is the density operator of the system it seeks to describe, in our case the cavity field. One way of facilitating the evaluation of this quantity is to invoke the theorem of Araki and Lieb \cite{araki} from which it is stated that if the composite atom-field system is initially in a pure state, then the entropies of the field and atomic subsystems would have to have equal values at $t>0$. So, instead of using the reduced density matrix of the field to evaluate its entropy, one can equivalently make use of the reduced operator associated with the atomic system alone (see Eq. (\ref{eq:rhoatoms})), for which, by virtue of its dimension, we just have a set of three eigenvalues $\lambda_{1,2,3}$, and then the field entropy acquires the simpler form 
\begin{equation}
S_{F} = -\sum_{i=1}^{3}\lambda_{i}\ln \lambda_{i},
\label{eq:entropyf}
\end{equation}
which ranges from 0, for a pure state, to $\ln (3)$, for a maximally mixed state. \\

\begin{figure}[h!]
\begin{center}
\includegraphics[width=7.6cm, height=5cm]{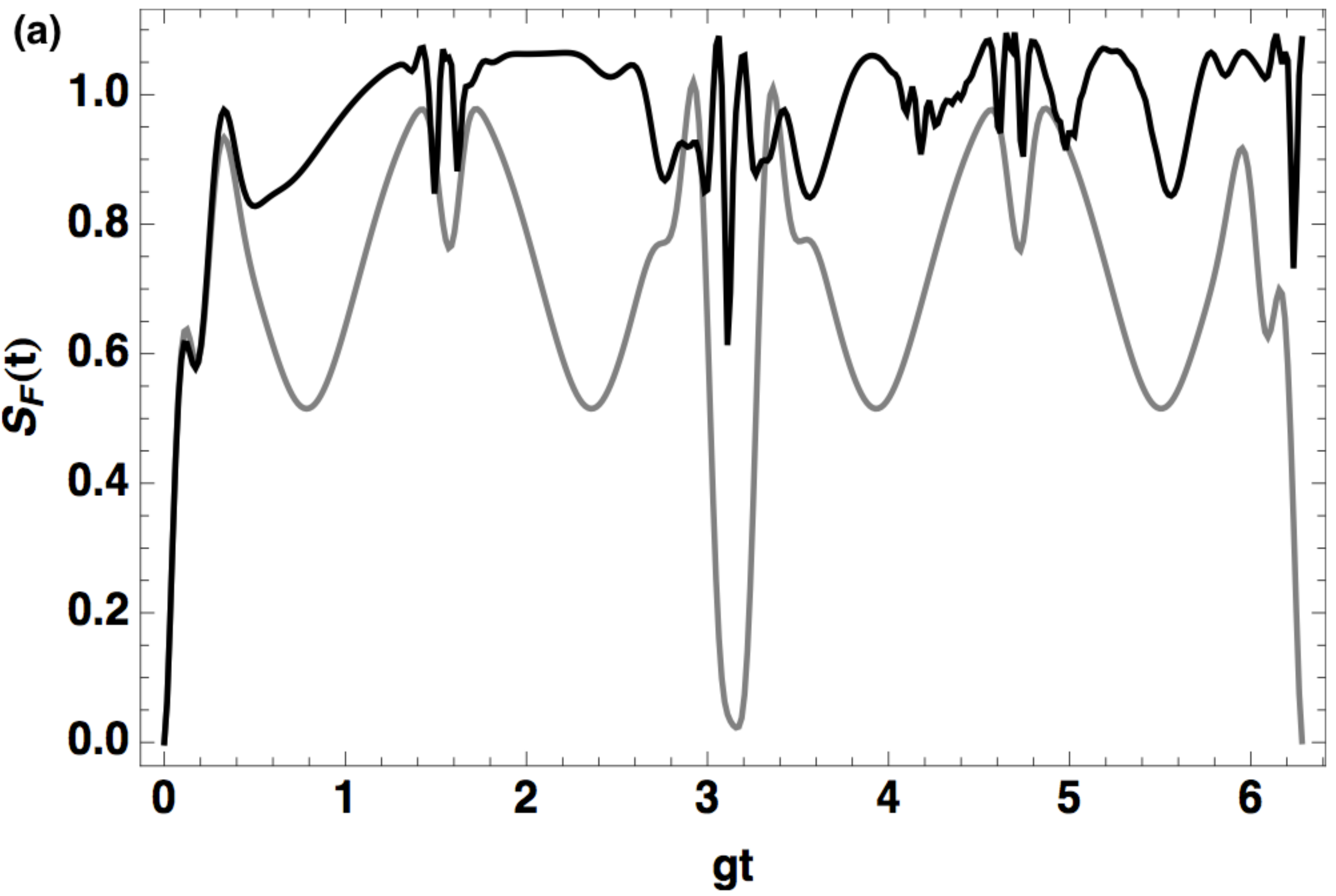} 
\includegraphics[width=7.6cm, height=5cm]{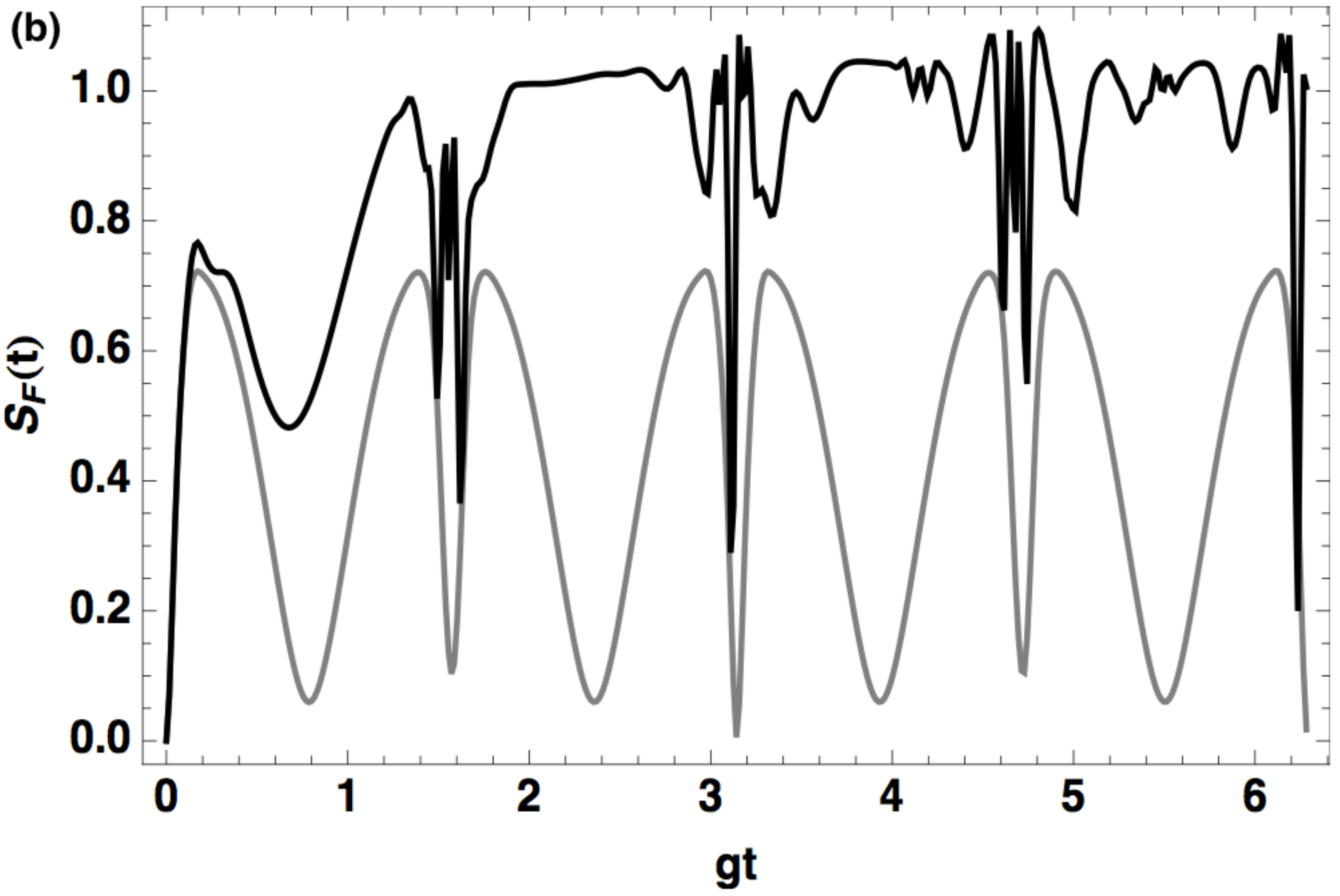} 
\caption{Plots of the evolution of field entropy calculated from Eq. (\ref{eq:entropyf}) when considering the atomic system being initially in its excited (frame (a)) and maximally entangled state (frame (b)) for a standard (gray curve) and a Kerr-like cavity (black curve). The field starts being a coherent state with $\langle \hat{n} \rangle=10$ photons and its coupling with the atoms is nonlinear in Buck-Sukumar's sense ($f(n)=\sqrt{n}$) and in the regime when $(\kappa-J)/g=0$.}
\label{fig:entropy1}
\end{center}
\end{figure}

In Fig. \ref{fig:entropy1} we show the evolution of the field entropy within the interval $0 \le gt \le 2 \pi$, where the field is initially in a coherent state with $\langle n \rangle =10$ photons, and the initial state of atomic system is such that the two atoms are considered to be in the excited state $|e,e\rangle$ (Fig. \ref{fig:entropy1} (a)) or the symmetric state $(|e,g\rangle+|g,e\rangle)/\sqrt{2}$ (Fig. \ref{fig:entropy1} (b)). In the figures, the black curves correspond to the case in which we have a linear coupling between the atoms and the field ($f(n)=1$), and the cavity is supposed to be filled with a Kerr-like medium such that $\chi/g=1/8$. The gray curves, on the other hand, describe the evolution of entropy when considering a Buck-Sukumar coupling, $f(n)=\sqrt{n}$, in the case of a standard cavity ($\chi=0$). Besides begin highly dependent upon the initial state of the atomic system, the general behavior of the field entropy seems unpredictable when both the linear atom-field coupling and the nonlinear medium are included, save for the appearance of slightly pronounced and barely regular spikes at certain time instants (see Fig. \ref{fig:entropy1} (b), black curve) indicating a tendency for the state of the field to recover, not completely, its purity when the atomic state is the maximally entangled one at the outset of its evolution. This last behavior is bolstered by the nonlinear Buck-Sukumar atom-field coupling within a standard cavity during the first stages of the time evolution (gray curves). During the shown time interval, there are specific time instants, integer multiples of $\pi/4$ in the dimensionless time variable $gt$, when the minima of the field entropy take place. At such time instants, the state of the field is almost pure when considering the state of the atoms to be initially in the maximally entangled state, a  behavior that seems to be also in agreement with the atomic purity in Fig. \ref{fig:purity-cs} (b) as expected. 

\begin{figure}[h!]
\begin{center}
\includegraphics[width=7.6cm, height=5cm]{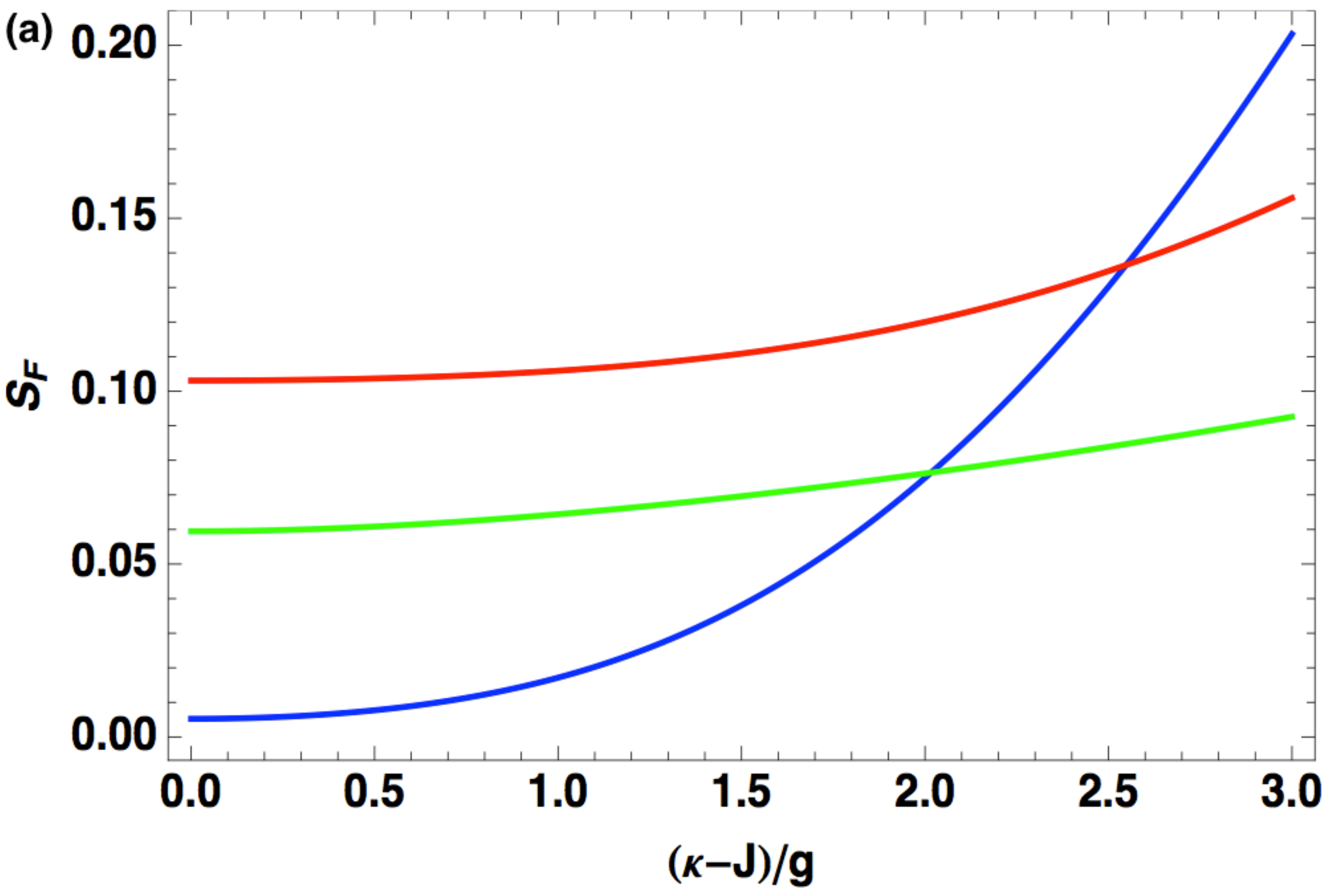} 
\includegraphics[width=7.6cm, height=5cm]{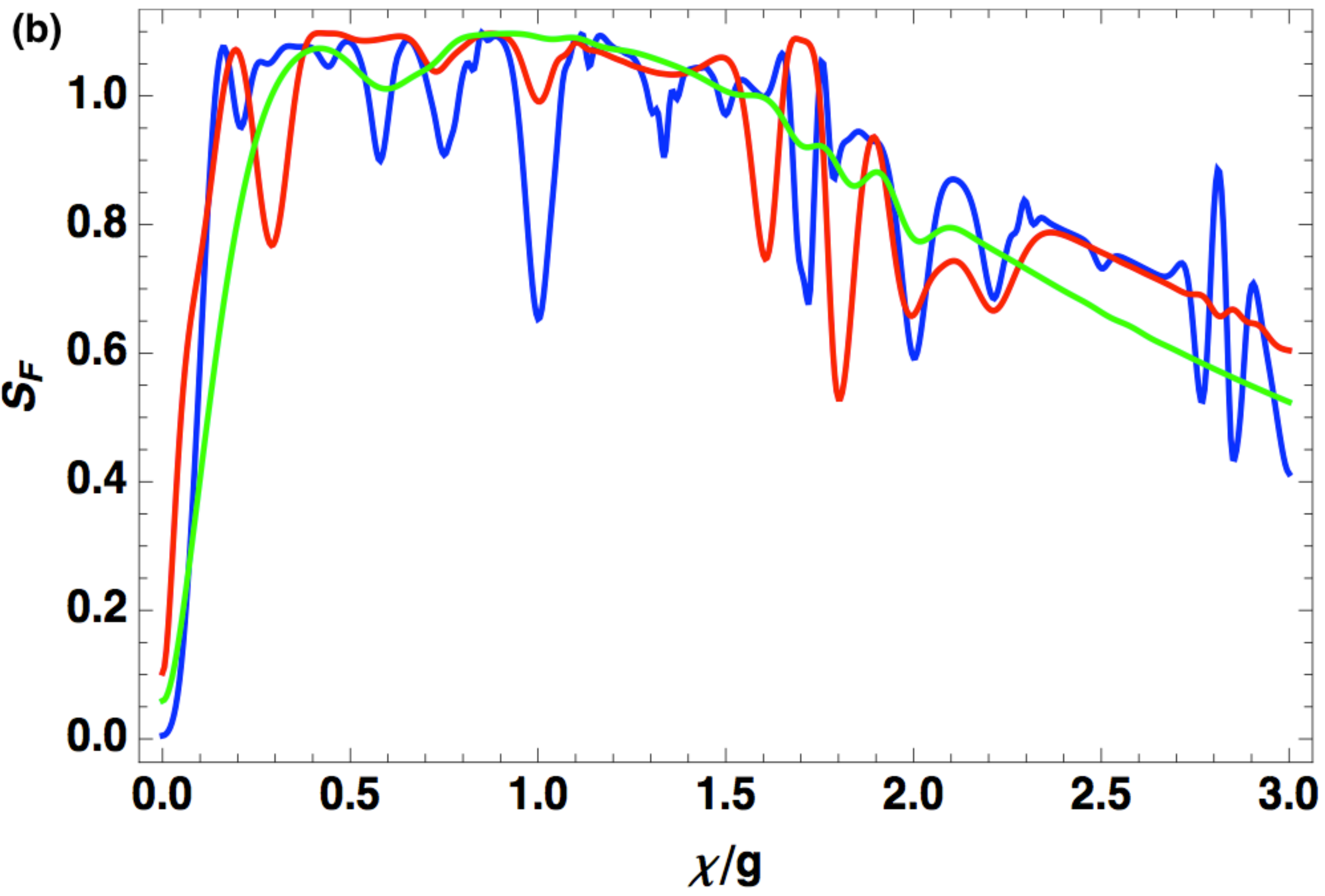} 
\caption{Field entropy calculated from Eq. (\ref{eq:entropyf}) and plotted, separately, as a function of $(\kappa-J)/g$ (frame (a)) and $\chi/g$ (frame (b)) for the time instants $gt=\pi/4$ (green), $\pi/2$ (red), and $\pi$ (blue). The field is initially in a coherent state with an average photon number $\langle \hat{n} \rangle=10$, and the interplay between it and the atoms takes place in the Buck-Sukumar regime ($f(n)=\sqrt{n}$). }
\label{fig:entropy2}
\end{center}
\end{figure}

As a function of the whole atomic and anharmonicity parameters, $(\kappa-J)/g$ and $\chi/g$, respectively, and in a wider range of their values than the one considered so far, it is depicted in Fig. \ref{fig:entropy2} the field entropy at the time instants $gt=\pi/4$, $\pi/2$ and $\pi$ for the nonlinear coupling case described in Fig. \ref{fig:entropy1} (b), gray curve, when the entropy seems to reach the first three minimum values. In similarity to the atomic purity, neither the dipole-dipole nor the Ising interaction has significant bearing on the entropy of the field, which means that such interactions preserve to some extent the purity of the subsystems. On the other hand, at the same time instants, the field evolves into a maximally mixed state quite rapidly with the increase of the anharmonicity parameter of the nonlinear medium, as seen in Fig. \ref{fig:entropy2} (b), within the interval $0\le \chi/g \le 0.5$, approximately. For $\chi/g >1.5$, we see once again a tendency to evolve into a pure state such as that of the atomic purity displayed in Fig. \ref{fig:purity-cs2} (b) in this strong nonlinear regime. \\

Let us now make a qualitative connection of the above results on entropy with a pictorial image provided by the so-called Q function. Before doing so, let us first establish the reduced density operator of the field computed via the following operation
\begin{equation}
\hat{\rho}_{F} = \Tr_{A} \{ \hat{\rho}_{AF} \},
\label{eq:tracerhoa}
\end{equation}
where $\Tr _{A}$ now indicates the trace over the atomic variables, and $\hat{\rho}_{AF}$, as stated earlier, denotes the density operator of the whole system composed of the field and the two atoms. Again, we shall regard the two atoms as being initially in their corresponding excited states, $|e,e\rangle$, or in the maximally entangled state, $(|e,g\rangle+|g,e\rangle)/\sqrt{2}$. So, starting from the density operator of the composite system given by Eq. (\ref{eq:densityaf}), together with the use of  (\ref{eq:tracerhoa}), the reduced density operator of the field looks like 
\begin{equation}
\hat{\rho}_{F} = \sum_{m,n=0}^{\infty} \sum_{k=1}^{3} A_{n}A_{m}^{\ast} \mathcal{D}_{k}^{(n)}(t)\mathcal{D}_{k}^{\ast (m)}(t)|n+k-1\rangle \langle m+k-1|,
\label{eq:rhof}
\end{equation}
and the time-dependent coefficient  $\mathcal{D}_{i}^{(n)}(t)$ is chosen to be either the one given by Eq. (\ref{eq:coefficient}) or Eq. (\ref{eq:coefficient2}), depending upon the initial state of the two atoms. 

By definition, the Husimi-Q distribution is given by
\begin{equation}
Q(\alpha) = \frac{1}{\pi} \langle \alpha |\hat{\rho}_{F} |\alpha \rangle,
\end{equation}
where, on substituting (\ref{eq:rhof}) into the last expression, we thus arrive at the desired result
\begin{equation}
Q(\alpha) = \frac{e^{-|\alpha|^{2}}}{\pi} \sum_{m,n}^{\infty} \sum_{k=1}^{3} \frac{A_{n}A_{m}^{\ast}\alpha^{m}\alpha^{\ast n}|\alpha|^{2(k-1)}\mathcal{D}_{k}^{(n)}(t)\mathcal{D}_{k}^{\ast (m)}(t)}{\sqrt{(m+k-1)!(n+k-1)!}}.
\label{eq:qfunctionf}
\end{equation}

So, with the help of the above expression for the Q function, some interesting features of the radiation field in phase space will be described by focusing our attention primarily upon the role played by the nonlinear character of the model. In what follows all the results concerning the evolution on phase space are presented by taking the cavity field to be initially in a coherent state with average photon number $\langle n \rangle =10$ and the atoms to be in the non-interacting regime, i.e., $\kappa-J=0$; this is so because the contribution of the atomic interplay to the field's evolution has proven so far to be subtle at least within the restricted range of values of the effective atomic parameter $(\kappa-J)/g$ and over the time scales considered here. \\

\begin{figure}[h!]
\begin{center}
\includegraphics[width=17cm, height=4cm]{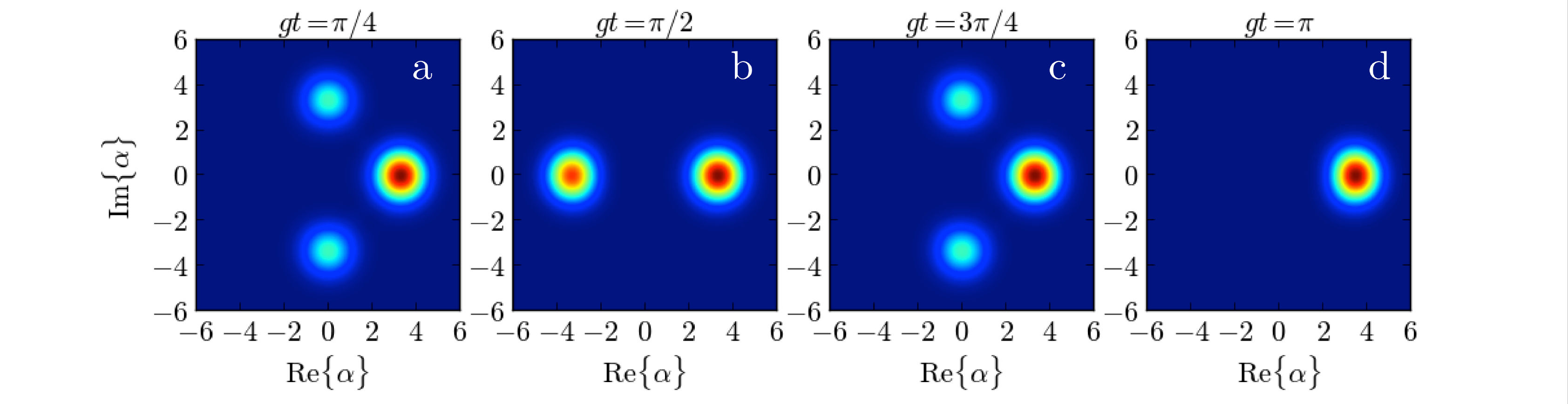} 
\includegraphics[width=17cm, height=4cm]{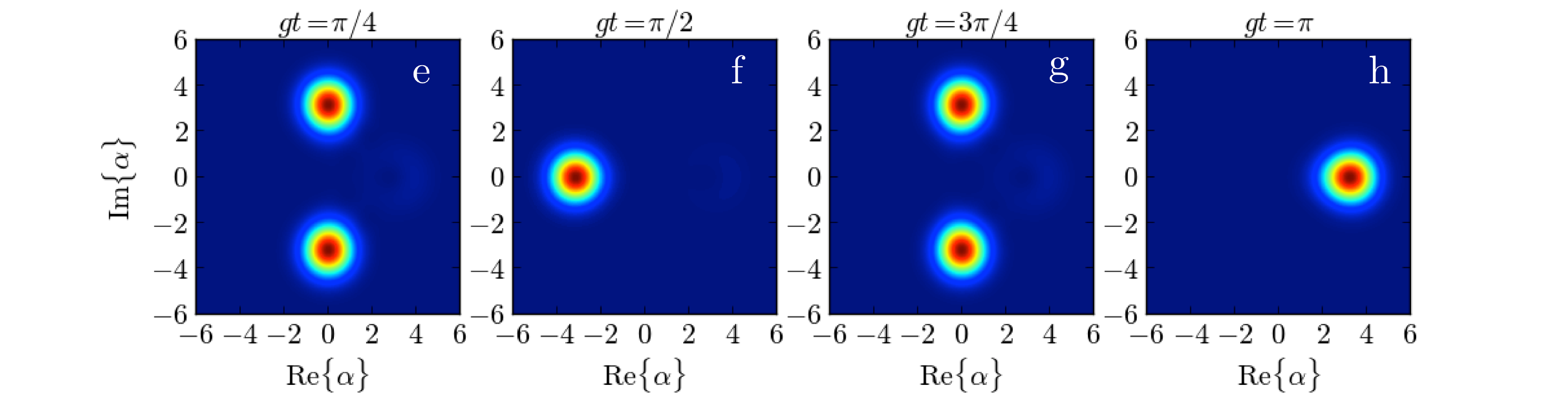} 
\caption{Contour plots of the Husimi-Q function for the field being initially in a coherent state with $\langle \hat{n} \rangle=10$ photons and the atomic system in its excited (upper row) or maximally entangled state (lowger row) for the time instants $gt=\pi/4,\pi/2,3\pi/4$, and $\pi$. Within the Buck-Sukumar regime, $f(n)=\sqrt{n}$, the plots are computed by considering the following parameters: $\delta=0$, $(\kappa-J)/g=0$ (non-interacting atoms), and $\chi/g=0$ (standard cavity).}
\label{fig:qfunction1}
\end{center}
\end{figure}

\begin{figure}[h!]
\begin{center}
\includegraphics[width=17cm, height=4cm]{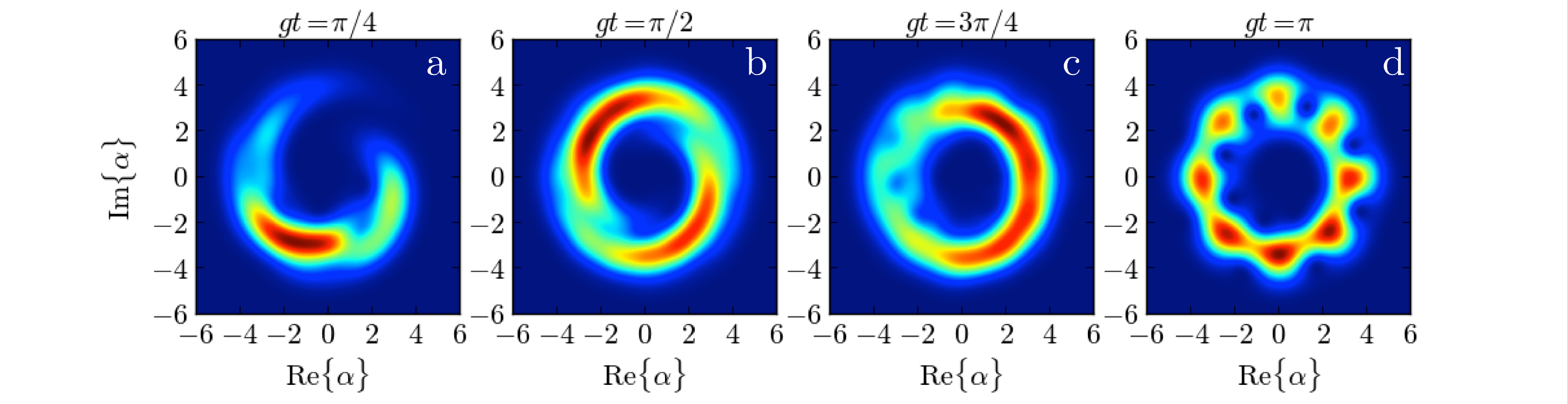} 
\includegraphics[width=17cm, height=4cm]{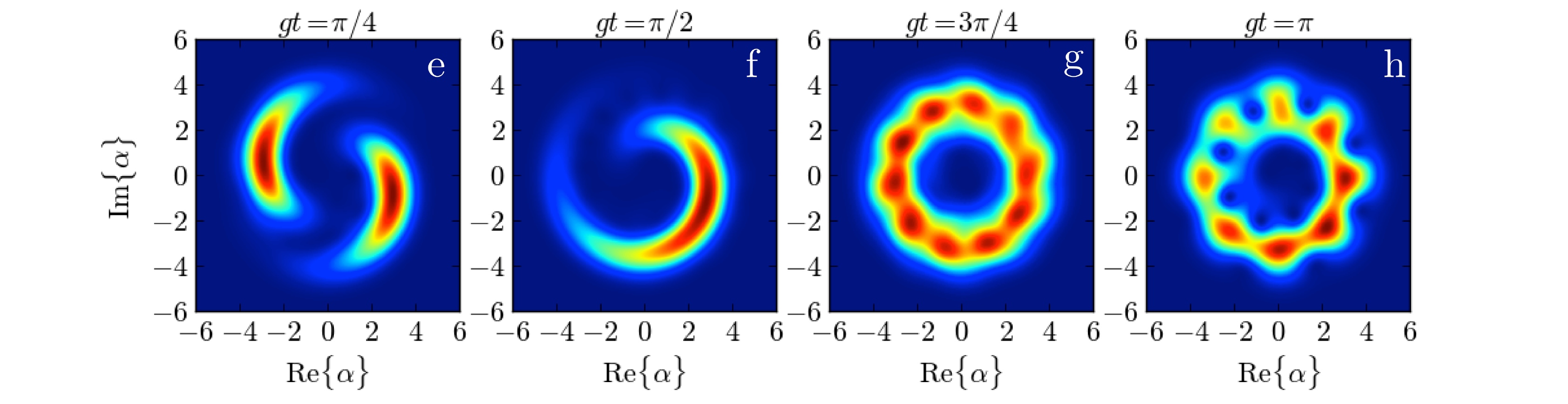} 
\caption{Contour plots of the Husimi-Q function for the field being initially in a coherent state with $\langle \hat{n} \rangle=10$ photons and the atomic system in its excited (upper row) or maximally entangled state (lowger row) for the time instants $gt=\pi/4,\pi/2,3\pi/4$, and $\pi$. Within the Buck-Sukumar regime, $f(n)=\sqrt{n}$, the plots are computed by considering the following parameters: $\delta=0$, $(\kappa-J)/g=0$ (non-interacting atoms), and $\chi/g=1/8$ (Kerr-like cavity).}
\label{fig:qfunction2}
\end{center}
\end{figure}

The corresponding results concerning the Buck-Sukumar regime, $f(n)=\sqrt{n}$, and the cases in which the field is allowed to evolve within a standard or a nonlinear Kerr-like cavity are shown in Figs \ref{fig:qfunction1} and \ref{fig:qfunction2}, respectively, at times $gt=\pi/4, \pi/2, 4\pi/3$ and $\pi$, just when the entropy of the field described earlier reaches approximately some of its first minimum values, which was particularly noticeable when a standard cavity is taken into consideration. In this regard, we  see in Fig. \ref{fig:qfunction1} that the development of the field turns out to be highly dependent upon the initial sate of the atomic system. For instance, if the two atoms are initially placed in their excited state (upper row), this circumstance can produce the formation of what seems to be well localized   phase components of the field, from two to three of them at times $gt=\pi/4$, $\pi/2$ and $3\pi/4$, that are reminiscent of the so-called Schr\"odinger-cat states, whereas only up to two phase components can be generated  when the atoms are initially in the maximally entangled state (lower row, at times $gt=\pi/4$ and $3\pi /4$). However, based on the results of the field's entropy (see Fig. \ref{fig:entropy1} (a), gray curve) at such time instants, neither the triple nor the double phase components are altogether pure states of the field, given that the minimum of entropy is not small enough to guarantee the purity of such states, thus having certain properties of a mixed state to some degree. This is not so when the field is almost re-focused into a single component, at  $gt=\pi$, as it started off at the outset of its evolution. On the other hand, if we now see Fig. \ref{fig:entropy1} (b), gray curve, as a reference, it is found that the state of the field characterized by the double phase components at $gt=\pi/4$ and $3\pi/4$, as well as the single ones at $gt=\pi/2$ and $\pi$, are almost pure states as far as the entropy's outcome is concerned. This conduct repeats itself periodically during the very first stages of field's evolution and its representation is constrained to develop around the phase space origin on a circle of radius $\sqrt{\langle n \rangle} \approx \sqrt{10}$.\\

As for the Kerr-like cavity, it is found that the nonlinear medium is able to create a richer cat-like splitting than the one we observe in the standard cavity case, as seen in some of the frames of Fig. \ref{fig:qfunction2}, say, at times $gt = 3\pi/4$ and/or $\pi$, depending on the initial state of the atomic system. But, according to entropy's outcome at such time instants (see, as a reference, Figs. \ref{fig:entropy1} (a) and (b), black curve), the degree of purity of the corresponding field states is not-so-significant as is the previous case; instead, most of the time they exhibit the proper character of a mixed state regardless of the initial state of the atoms, and their frequency of appearance is no longer periodic as a function of time.  At the beginning of the evolution, the Kerr effect manifests itself in the distortion of the field, namely, in some cases, the phase of it is warped and curved in a single or double deformed elongated rim at first (see, for instance, frame (a), upper row; frames (e) and (f), lower row) and then is spread over a kind of phase space ring composed of a number of cat-like states as time elapses (frame (g), lower row). This behavior was proven to be more unpredictable throughout an extended time interval far beyond the one pointed out in the present results. 

Finally, it is worth mentioning that the Kerr effect on field's phase space has thoroughly been investigated by several authors in several circumstances involving atom-field interactions, among them we can cite the work of Werner and Risken \cite{werner} on the dynamics of the Q-function for a two-level atom; the generation of superpositions of distinguishable states by means of  Yurke and Stoler's model \cite{yurke}; the treatment of one- and two-photon Jaynes-Cummings models with one atom via a Pegg-Barnett Hermitian phase operator formalism undertaken by Gantsog \etal \cite{tanas}; and, more recently, a nonlinear version of the Jaynes-Cummings model and its nonlinear coherent states based upon the f-deformed oscillator formalism has also been introduced \cite{octa}.


\section{Conclusions}

In this work we have put forward and provided the solution of a nonlinear version of the Jaynes-Cummings model for two identical two-level atoms allowing for Ising-like and dipole-dipole interactions between them. The model is nonlinear in a twofold sense, that is, it incorporates the possibility of describing the evolution of the radiation field in a cavity that can be supposed to be filled with an nonlinear medium and/or a general intensity-dependent coupling between the field and the two atoms, each of these characteristics being specified, separately, by the proper photon-number-dependent function. Even though we have considered two two-level atoms interacting (nonlinearly) with a quantized field, we emphasize the fact that such nonlinearities may appear in analogous systems (in the sense that Hamiltonians like the one we consider here may be engineered) such as trapped ions interacting with classical laser fields \cite{wallentowitz,matos,manko,blatt,moya}. If, for instance, we consider two trapped ions, close enough together to interact between them, as represented in (\ref{eq:haa}), and a set of lasers detuned correctly, one can generate nonlinearities that behave as associated Laguerre polynomials, and by adding properly, one may easily reproduce a number of possible nonlinear models we are familiar with from nonlinear optics,  such as the usual Kerr-type and Buck-Sukumar nonlinearities or even more complicated coupling schemes. Therefore, our treatment is not only valid for the interaction between a quantized field and two two-level atoms, but also for other (algebraically) similar systems. \\

Some dynamical properties of the atomic and field subsystems, such as the atomic excitation, purity and concurrence, as well as the entropy and phase space dynamics of the field, have been described by considering a Kerr-like cavity and/or a Buck-Sukumar atom-field coupling, as particular examples, in the resonant quantum dynamics ($\delta=0$). The most interesting outcomes of the model come from featuring such nonlinearities or merging any of them with the atomic interplay under certain circumstances. In this regard, it was found that the combination of the nonlinear atom-field coupling and the dipole-dipole-and-Ising-like atomic interaction, in the range of values $0<(\kappa-J)/g \le1/2$, or merging the former with the anharmonic effect of a Kerr-type cavity, within the regime $0<\chi/g\ll 1$, gives rise to the appearance of a noticeable beating pattern and the inversion of successive revivals in the evolution of atomic excitation, and from the algebraic results we have been able to obtain an approximate closed expression for the said property that has allowed us to estimate the origin and duration of such beats, as well as the collapse and revival times. Both the purity and concurrence atomic properties are bolstered by the Buck-Sukumar coupling itself at certain time instants, and by the Kerr-like medium up to some degree, particularly when the atomic system is in its maximally entangled state. As to the atomic purity and field entropy, it was also found that the dipole-dipole and Ising effects, though not-so-significant by themselves, become manifest in the regime when the ratio of the whole atomic parameter to the constant coupling is such that $(\kappa-J)/g \gg 1$, but nevertheless we suggest that one should be careful of working on this regime given that, as far as we know, it has not been proven in the literature to be suitably justified at such high values; this is why we have chosen here to be moderate in the use of small values of the aforesaid parameter. Finally, we have explored the dynamics of the field in terms of its entropy and evolution on phase space via the Q-function, where the role played by Buck-Sukumar nonlinear coupling and the Kerr medium were found to be preponderant in the splitting of  phase properties of the field giving rise to Schr\"odinger-cat-like states.

\ack We aknowledge partial support from projects DGAPA UNAM IN113016 and CONACyT 166961. One of the authors (O de los Santos-S\'anchez) wishes to thank the {\it Red Tem\'atica de Tecnolog\'{\i}as Cu\'anticas} for financial support through project 250785 CONACyT-CNPq M\'{e}xico. CGG is grateful to CONACyT for financial support under doctoral fellowship No. 385108.
Last but not least, we especially thank both professor H. M. Moya-Cessa and an anonymous referee for helpful suggestions that greatly improved the content of the paper.

\end{document}